\documentclass{article}[jheppub]
\usepackage{pdfpages}
\usepackage{float}
\usepackage{amsmath}
\usepackage{mathtools}
\usepackage{amsfonts}
\usepackage{bm}
\usepackage{jheppub}
\usepackage{color}
\usepackage{subfigure}
\usepackage[utf8x]{inputenc}
\newcommand{\beq}{\begin{equation}}
\newcommand{\eeq}{\end{equation}}
\newcommand{\beqn}{\begin{eqnarray}}
\newcommand{\eeqn}{\end{eqnarray}}
\newcommand\be{\begin{equation}}
\newcommand\ee{\end{equation}}
\newcommand\bea{\begin{eqnarray}}
\newcommand\eea{\end{eqnarray}}
\newcommand\Dzerobar{\overline{D}_0}
\newcommand\Czerobar{\overline{C}_0}
\newcommand\Bzerobar{\overline{B}_0}

\def\three{{\bm{3}}}
\def\four{{\bm{4}}}
\def\five{{\bm{5}}}

\def\mC{{\cal C}}
\def\mCbar{{\cal \bar{C}}}
\def\slsh{\rlap{$\;\!\!\not$}}     
\def\e{\epsilon}
\def\cG{r_\Gamma}
\def\Higgs{H}
\def\mh{M_\Higgs}
\def\tr{{\rm tr}}

\def\qb{\bar{q}}
\newcommand{\x}{\times}
\def\mC{\mathcal{C}}

\def\spa#1.#2{\left\langle#1\,#2\right\rangle}
\def\s#1.#2{s_{#1#2}}
\def\spb#1.#2{\left[#1\,#2\right]}
\def\spab#1.#2.#3{\left\langle#1|#2|#3\right]}
\def\spba#1.#2.#3{\left[#1|#2|#3\right\rangle}
\def\spaa#1.#2.#3.#4{\left\langle#1|#2|#3|#4\right\rangle}
\def\spbb#1.#2.#3.#4{\left[#1|#2|#3|#4\right]}
\def\spaabb#1.#2.#3.#4.#5{\left\langle#1|#2|#3|#4|#5\right]}
\def\spaba#1.#2.#3.#4{\left\langle#1|#2|#3|#4\right\rangle}
\def\spbab#1.#2.#3.#4{\left[#1|#2|#3|#4\right]}
\def\spbaba#1.#2.#3.#4.#5{\left[#1|#2|#3|#4|#5\right\rangle}
\def\spabab#1.#2.#3.#4.#5{\left\langle#1|#2|#3|#4|#5\right]}
\def\spbabab#1.#2.#3.#4.#5.#6{\left[#1|#2|#3|#4|#5|#6\right]}
\def\spababa#1.#2.#3.#4.#5.#6{\left\langle#1|#2|#3|#4|#5|#6\right\rangle}
\def\trfive{{\rm tr}_{5}}
\def\TrfourgL#1.#2.#3.#4{{\rm tr}_{-}\{{#1}\,{#2}\,{#3}\,{#4}\}}
\def\TrfourgR#1.#2.#3.#4{{\rm tr}_{+}\{{#1}\,{#2}\,{#3}\,{#4}\}}
\def\Trfour#1.#2.#3.#4{{\rm tr}\{{#1}\,{#2}\,{#3}\,{#4}\}}

\def\DeltaThree{\Delta_3}

\def\be{\begin{equation}}
\def\ee{\end{equation}}
\def\bea{\begin{eqnarray}}
\def\eea{\end{eqnarray}}
\def\beal{\begin{equation}\begin{aligned}}
\def\eeal{\end{aligned}\end{equation}}
\def\nn{\nonumber}

\def\Res_#1{\operatorname*{Res}_{#1}}

\def\tr{\operatorname*{tr}}

\author[a]{John M. Campbell,}
\emailAdd{johnmc@fnal.gov}
\author[b]{Giuseppe De Laurentis,}
\emailAdd{giuseppe.delaurentis@ed.ac.uk}
\author[c]{R. Keith Ellis,}
\emailAdd{keith.ellis@durham.ac.uk}

\affiliation[a]{Fermilab, PO Box 500, Batavia IL 60510-5011, USA}
\affiliation[b]{Higgs Centre for Theoretical Physics, University of Edinburgh, Edinburgh, EH9 3FD, UK}
\affiliation[c]{Institute for Particle Physics Phenomenology, Durham University, Durham, DH1 3LE, UK}

\preprint{FERMILAB-PUB-24-0498-T,\, IPPP/24/56}
\title{Analytic amplitudes for a pair of Higgs bosons in association with three partons}

\abstract{The pair production of Higgs bosons at the LHC can give
  information about the triple Higgs boson coupling. We perform an
  analytic one-loop calculation of the amplitudes for a pair of Higgs
  bosons in association with three partons, retaining the exact
  dependence on the quark mass circulating in the loop.  These
  amplitudes constitute the real radiation corrections in the
  calculation of Higgs boson pair production at next-to-leading order
  in the strong coupling. The results of an analytic
  generalised-unitarity computation are simplified via analytic
  reconstruction in spinor variables. Compact ans\"atze for kinematic
  pole residues are iteratively fitted via $p$-adic evaluations near
  said poles and subtracted until no pole remains. A new ansatz
  construction is introduced to minimally parametrise coefficients of
  amplitudes with multiple massive external legs. The simplified
  expressions are faster to evaluate than automatic codes and can lead
  to more stable results near singular regions.}

\begin{document}
\maketitle

\section{Introduction}
The exploration of the details of the Higgs phenomenon is one of the
primary goals of the high-luminosity LHC and its successor machines.
The exploration of the Higgs potential is an important
part of that endeavour.
Within the context of the standard model, where the Lagrangian
is limited to terms of mass dimension $d\leq4$, the Higgs potential is
fully determined,
\beq \label{Vpot}
V(h)=\frac{1}{2} \mh^2 H^2 +\lambda v H^3+\frac{1}{4} \lambda H^4 \, ,
\eeq
in terms of the Higgs boson mass, $\mh$, and the Fermi constant, $G_F$,
where $\mh=\sqrt{2 \lambda} \, v$, and $G_F/\sqrt{2}=1/(2 v^2)$.
Beyond the standard model one can introduce operators of dimension
higher than four, and hence deviations from the simple form for
the triple Higgs boson coupling given in eq.~(\ref{Vpot}).
For a complete review we refer the reader to ref.~\cite{DiMicco:2019ngk}.

Non-perturbative analyses~\cite{Csikor:1998eu}
indicate that, at the observed mass of the Higgs boson, the electroweak phase transition
is a rapid crossover, but extensions of the standard model can make it first order.
A first order phase transition would be required, inter alia, for an electroweak baryogenesis
explanation of the baryon asymmetry of the universe. This explains the intense interest in extending
our knowledge of the Higgs potential, beyond the limited information about the shape of the potential
derived from quadratic excursions about the minimum of the potential, governed by the Higgs boson mass.

Constraints on the triple Higgs boson coupling derive from measurements
of both single Higgs~\cite{McCullough:2013rea} and
double Higgs boson production~\cite{CMS:2018ipl,CMS:2022dwd,ATLAS:2019qdc,ATLAS:2022jtk}.
The production of Higgs boson pairs gives direct access to the self-coupling of the Higgs boson.
Deviations from the standard model are most easily assessed in the kappa
framework~\cite{LHCHiggsCrossSectionWorkingGroup:2013rie,LHCHiggsCrossSectionWorkingGroup:2016ypw},
in which the standard model triple Higgs boson coupling
is allowed to float by an overall factor $\kappa_\lambda$.
In ref.~\cite{ATLAS:2024lhu} the ATLAS collaboration find limits on the
triple Higgs boson coupling modification of $-6.3 < \kappa_\lambda < 11.6$ at $95\%$ CL,
using final states with leptons (including taus), and photons.
A combination limit from ATLAS in ref.~\cite{ATLAS:2024ish} using final states
including $b$-quarks, determines that $\kappa_\lambda$
lies in the range $-1.2 < \kappa_\lambda < 7.2 $ at $95\%$ CL.
In ref.~\cite{CMS:2024awa} the CMS collaboration exclude values of the coupling modifier
outside the range $-1.2 < \kappa_\lambda < 7.5 $ at $95\%$ CL,
using results from both Higgs boson pair production and single
Higgs boson production.

These limits on the triple Higgs boson coupling rely on theoretical calculations of
Higgs boson processes at next-to-leading order (NLO) and beyond.
NLO corrections to Higgs boson pair production including full top quark mass dependence have been calculated in
refs.~\cite{Borowka:2016ehy,Borowka:2016ypz,Baglio:2018lrj,Baglio:2020ini}.
These calculations have been used to introduce top quark mass dependence into NNLO
calculations performed in the $m\to \infty$ limit in refs.~\cite{Grazzini:2018bsd,Chen:2019lzz,Chen:2019fhs}.
The theoretical uncertainties due to renormalization and factorization scale choice, with a special focus
on the renormalization scheme for the top quark mass
have been discussed in ref.~\cite{Baglio:2020wgt}.
The effects of matching with a parton shower are described in refs.~\cite{Heinrich:2017kxx,Bagnaschi:2023rbx}.
The calculation of double Higgs boson production at NLO has been further improved by combining
the numerical fit result,
based on events with limited coverage of the high-energy region, with
the high-energy expansion, yielding a NLO result valid in the low, medium and high-energy regions~\cite{Davies:2019dfy}.
For a short recent review including Higgs boson pair production we refer the reader to ref.~\cite{Jones:2023uzh}.

In this paper we re-examine part of the NLO calculation of Higgs boson pair production, namely the one-loop amplitudes
for the processes,
\beqn
0 &\to& g(p_1) +g(p_2) +g(p_3) +H(p_4)+H(p_5) \,,\\
0 &\to& q(p_1) +\qb(p_2) +g(p_3) +H(p_4)+H(p_5) \,,
\eeqn
retaining all the dependence on the top quark mass, $m$.
All the results presented assume that the two Higgs bosons are both on their mass shell.
These processes represent the real radiation contribution to the NLO Higgs boson pair production process.
The motivation for doing this is twofold. First, we hope to produce
expressions which are faster to evaluate than expressions generated
with automatic tools\footnote{In NLO corrections the two-loop virtual correction
is represented by a fit~\cite{Heinrich:2017kxx}, which is quite fast to evaluate. In previous calculations the one-loop
real contribution is responsible for a sizable part of the computation time.}. Second, in NLO calculations  the real radiation
is probed in regions where the emitted parton is either soft or collinear
with respect to the initial partons and the one-loop expressions can be quite unstable.
We achieve these goals by simplifying the spinor expressions for our results
using the techniques of refs.~\cite{DeLaurentis:2019bjh,DeLaurentis:2022otd}. This has benefits
for the evaluation time, since the expressions for the coefficients are
shorter, and also for the stability of the results. This is the case since
rational functions are for the most part reduced to least common denominator
form, and partial fraction decompositions are used to reveal their
underlying divergence structure.

In addition to presenting the analytic forms for the amplitudes, we perform a
comparison with matrix elements calculated using automatic procedures.
By reusing the prior calculation of the $ggHH$ 2-loop contribution~\cite{Heinrich:2017kxx}, we
also present a new public implementation of the NLO Higgs-pair production
process in {\tt MCFM}~\cite{MCFM,Campbell:2015qma,Campbell:2019dru}.  Additionally, this calculation is used to
provide a result that is accurate to NLO+NNLL at small transverse momentum of
the Higgs boson pair, which may be of interest at a future high-energy
$pp$ collider.

\section{One-loop amplitude for $gg HH$}
We first review the amplitude for the lowest order Higgs boson pair production process,
which at order $\alpha_s$ in the strong coupling, occurs through the one-loop process,
\beq
\label{lowestorderamp}
  0 \to g(p_1)+g(p_2)+H(p_3)+H(p_4)\, .
\eeq
We present these results here for completeness and to introduce our notation.
The full one-loop amplitude, known for many years, is given in ref.~\cite{Glover:1987nx}.
This result supersedes an earlier partial result in ref.~\cite{Eboli:1987dy}.
Since much of the interest in this process 
derives from its sensitivity to the trilinear coupling of the Higgs boson,
we modify that term in the Lagrangian by introducing a rescaling $\kappa_\lambda$,
\beq
    {\cal L}_{\rm Higgs}=\frac{1}{2} \partial_\mu H\partial^\mu H -\frac{1}{2} \mh^2 H^2 -\kappa_\lambda \lambda v H^3 + \ldots\, .
\eeq
We have dropped the quartic coupling of the Higgs boson since it plays no part in the present calculation.
The full amplitude for the process in eq.~(\ref{lowestorderamp}) is given by,
\begin{equation}
  -i {\cal{A}}^{C_1 C_2}=-\frac{1}{2}\delta^{C_1 C_2} \, \alpha_s \, \frac{\alpha_W }{4 M_W^2}\, A(1_g,2_g,3_H,4_H)\,,
\end{equation}
where,
\begin{equation}
  M_W=\frac{1}{2}g_W v,\;\;\alpha_s=\frac{g_s^2}{4 \pi}\, ,\;\;\alpha_W=\frac{g_W^2}{4 \pi}\,.
\end{equation}
$C_1$ and $C_2$ are the colour indices of the gluons, $v\simeq 246$~GeV is the vacuum expectation value of the Higgs field, $M_W$ is the mass of the $W$-boson,
$g_W$ is the gauge coupling of the $SU(2)_W$ weak gauge group, and $g_s$ is the gauge coupling of the $SU(3)$ strong gauge group. 
There are two independent helicity amplitudes,
\begin{eqnarray} \label{LOhelicities}
  A(1_g^+,2_g^+,3_H,4_H)&=& (\kappa_\lambda \, g_1^{\triangle}+g_1^{\square}) \frac{\spb1.2^2}{s_{12}} \,, \nn \\
  A(1_g^+,2_g^-,3_H,4_H)&=& g^{\square}_2 \frac{\spba1.\three.2^2}{s_{12} p_T^2}\,,
\end{eqnarray}
where we have defined,
\beqn
&&s_{12}=(p_1+p_2)^2,\;\;
\mh^2=p_3^2=p_4^2\equiv(p_1+p_2+p_3)^2,\;\; \nn \\
&&\frac{2 p_1 \cdot p_3\, p_2 \cdot p_3}{p_1\cdot p_2}=p_T^2+\mh^2\,,
\eeqn
and $\mh$ is the Higgs boson mass.
The remaining two helicity combinations are obtained from eq.~(\ref{LOhelicities}) by interchange.
$ g_1^{\triangle}$ denotes the triangle-graph pieces of the amplitude, contributing via the triple Higgs boson coupling.
The determination of the triple Higgs boson coupling provides much of the
motivation for the experimental measurement of Higgs boson pair production. 
In this equation $\spb{i}.{j}$ and $\spab{i}.{\bm k}.{j}$ are Lorentz invariant contractions of the spinors with momentum $p_i,p_j$ and $p_k$. 
The momenta $p_i$ and $p_j$ are lightlike, whereas $p_k^2 =\mh^2$. As a reminder of its non-zero mass, in spinor products massive four-vectors are written in boldface.
\begin{equation}
\label{Spinor_products_simple}
\spa i.j=\bar{u}_-(p_i) u_+(p_j), \;\;\;
\spb i.j=\bar{u}_+(p_i) u_-(p_j), \;\;\;
\spab i.{\bm k}.j =\bar{u}_-(p_i)\slsh{k} u_-(p_j)\,.
\end{equation}
Full details of the spinor notation and more complicated spinor strings, such as $\spaba1.\four.\five.2$ are given in Appendix~\ref{spinorsection}.

The results for the coefficients $g_1^\triangle,g_1^\square$ and $g_2^\square$ are~\cite{Glover:1987nx},
\begin{eqnarray}
  g_1^\triangle&=& \frac{12 m^2 \mh^2}{s_{12}-\mh^2} \Big[ 2 +(4m^2-s_{12})C_0(p_1,p_2) \Big] \,,\\
  g_1^\square &=& 4 m^2 \bigg\{ m^2 (8 m^2-s_{12} -2 \mh^2) \big(D_0(p_1,p_2,p_3;m)+D_0(p_2,p_1,p_3;m)+D_0(p_1,p_3,p_2;m)\big) \nn \\
  &+&\frac{(s_{1\three} s_{2\three} -\mh^4)}{s_{12}}(4 m^2-\mh^2) D_0(p_1,p_3,p_2;m)+2+4m^2 C_0(p_1,p_2;m)\nn \\
  &+& \frac{2}{s_{12}} (\mh^2-4m^2)\big((s_{1\three}-\mh^2)C_0(p_1,p_3;m)+(s_{2\three}-\mh^2)C_0(p_2,p_3;m)\big)\bigg\} \,,
\end{eqnarray}
\begin{eqnarray}
  g_2^\square &=&2 m^2 \bigg\{ 2 (8 m^2+s_{12} -2 \mh^2) \nn \\
  &\times &\big\{m^2 [D_0(p_1,p_2,p_3;m)+D_0(p_2,p_1,p_3;m)+D_0(p_1,p_3,p_2;m)]-C_0(p_3,p_4;m)\big\} \nn \\
  &-&2 \big\{ s_{12} C_0(p_1,p_2;m)+(s_{1\three}-\mh^2) C_0(p_1,p_3;m)+(s_{2\three}-\mh^2) C_0(p_2,p_3;m)\big\}\nn \\
  &+& \frac{1}{(s_{1\three} s_{2\three}-\mh^4)} \bigg[s_{12} s_{2\three} (8 s_{2\three} m^2-s_{2\three}^2-\mh^4)D_0(p_1,p_2,p_3;m)\nn\\
  &+& s_{12} s_{1\three} (8 s_{1\three} m^2-s_{1\three}^2-\mh^4) D_0(p_2,p_1,p_3;m)\nn \\
    &+& (8m^2+s_{12}-2 \mh^2) \big\{s_{12} (s_{12}-2\mh^2)C_0(p_1,p_2;m)+s_{12}(s_{12}-4 \mh^2)C_0(p_3,p_4;m)\nn \\
    &+&2 s_{1\three} (\mh^2-s_{1\three})C_0(p_1,p_3;m)+2 s_{2\three} (\mh^2-s_{2\three})C_0(p_2,p_3;m)\big\}\bigg]\Bigg\}\,,
\end{eqnarray}
with
\beqn
s_{1\three}=(p_1+p_3)^2,\;\;
s_{2\three}=(p_2+p_3)^2,
\eeqn
and where $m$ is the (top) quark mass.
$B_0,C_0$ and $D_0$ are bubble, triangle and box scalar integrals respectively in a more-or-less
standard notation~\cite{Passarino:1978jh}. Full details of the notation for scalar integrals are given in Appendix~\ref{Integrals}.
To emphasize that the momenta of the Higgs bosons $p_3$ (and $p_4$)
are not light-like, we denote their presence in scalar products in boldface, thus, for example, $s_{1\three}=(p_1+p_3)^2$.

\subsection{General decompositions of one-loop amplitudes}
\label{sec:decomposition}
The one-loop amplitude for $ggHH$ was naturally expressed in terms of (tadpole), bubble, triangle and box scalar integrals,
because the amplitude had four external lines.
Scalar integrals are loop integrals with no powers of the loop momentum in the numerator.
The tadpole integral is not needed since it can be eliminated in terms of a bubble integral and a rational term.
However it is more generally true that one loop amplitudes can be expressed as a sum of bubble, triangle and box scalar integrals.
Thus even in the case of higher point amplitudes with a larger number of external legs it is still true that we may write,
\begin{eqnarray} \label{generaldecomposition}
  A & = & \frac{\bar\mu^{4-n}}{r_\Gamma}\frac{1}{i \pi^{n/2}} \int {\rm d}^n \ell
 \, \frac{{\rm Num}(\ell)}{\prod_i d_i(\ell)} \nn \\
&=& \sum_{i,j,k} {d}_{i\x j\x k}(1^{h_1},2^{h_2},3^{h_3}) \, D_0(p_i, p_j, p_k ;m)  \nn \\
&+& \sum_{i,j} {c}_{i\x j}(1^{h_1},2^{h_2},3^{h_3}) \,  C_0(p_i,p_j ;m)   \nn \\
&+& \sum_{i} {b}_{i}(1^{h_1},2^{h_2},3^{h_3}) \, B_0(p_i;m) + r(1^{h_1},2^{h_2},3^{h_3})\, .
\end{eqnarray}
The scalar bubble ($B_0$), triangle ($C_0$), box ($D_0$) integrals, and the
constant $r_\Gamma$, are defined in Appendix~\ref{Integrals}. This decomposition has the merit
that the number of loop integrals that need to be evaluated is minimized.

For the case of the amplitude $gggHH$ we have pentagon diagrams with 5 external legs, but the
general decomposition in Eq.~(\ref{generaldecomposition}) still holds. Indeed in four dimensions,
the scalar pentagon integral, $E_{0}$ is expressible as a sum of 5 scalar boxes obtained by removing
the denominators of $E_{0}$ one at a time,
\begin{eqnarray} \label{pentagonreduction}
E_{0}(p_1,p_2,p_3,p_4;m)&=&
 \mC^{1\x2\x3\x4}_1\,D_0(p_2,p_3,p_4;m)
+\mC^{1\x2\x3\x4}_2\,D_0(p_{12},p_3,p_4;m) \nn \\
&+&\mC^{1\x2\x3\x4}_3\,D_0(p_1,p_{23},p_4;m) \nn \\
&+&\mC^{1\x2\x3\x4}_4\,D_0(p_1,p_2,p_{34};m)
+\mC^{1\x2\x3\x4}_5\,D_0(p_1,p_2,p_3;m)\, .
\end{eqnarray}
Rules for calculating the coefficients $\mC$ are given in section~\ref{sec:pentbox}.
Even though Eq.~(\ref{generaldecomposition}) is always true, we find that for the $gggHH$ process
there is merit in using a more general decomposition that retains the pentagon integrals 
and yields more compact results for the coefficients of the integrals.  In this basis we have,
\begin{eqnarray} \label{Pentagondecomposition}
  A & = & \frac{\bar\mu^{4-n}}{r_\Gamma}\frac{1}{i \pi^{n/2}} \int {\rm d}^n \ell
 \, \frac{{\rm Num}(\ell)}{\prod_i d_i(\ell)} \nn \\
&=& \sum_{i,j,k,l} {\hat{e}}_{i\x j\x k \x l}(1^{h_1},2^{h_2},3^{h_3}) \, E_0(p_i, p_j, p_k, p_l ;m)  \nn \\
&+& \sum_{i,j,k} {\hat{d}}_{i\x j\x k}(1^{h_1},2^{h_2},3^{h_3}) \, D_0(p_i, p_j, p_k ;m)  \nn \\
&+& \sum_{i,j} {c}_{i\x j}(1^{h_1},2^{h_2},3^{h_3}) \,  C_0(p_i,p_j ;m)   \nn \\
&+& \sum_{i} {b}_{i}(1^{h_1},2^{h_2},3^{h_3}) \, B_0(p_i;m) + r(1^{h_1},2^{h_2},3^{h_3})\, .
\end{eqnarray}
In our final results, expressions for the box coefficients ${d}_{i\x j\x k}$ in Eq.~(\ref{generaldecomposition})
are given in terms of the appropriate sum of effective pentagon and remainder box coeffients,
${\hat{e}}_{i\x j\x k \x l}$ and ${\hat{d}}_{i\x j\x k}$ respectively,
in Eq.~(\ref{Pentagondecomposition}).
This method of effective pentagons has also been used in the description of the $Hgggg$ process
in ref.~\cite{Budge:2020oyl}.

\section{Advancements in analytic reconstruction techniques}
\label{sec:advancements}

The one-loop coefficients contributing to the process with an
additional parton in the final state, $pp\rightarrow HHj$, are
presented here in the form obtained through analytic
reconstruction. They are iteratively reconstructed one pole residue at
a time, as described in ref.~\cite[Section 3.3]{DeLaurentis:2019bjh}.
The use of algebraic geometry ensures control over the analytic
structure of the coefficients, while $p\kern0.2mm$-adic numbers enable
stable numerical evaluations \cite{DeLaurentis:2022otd}. See also
related work in ref.~\cite{Campbell:2022qpq, Chawdhry:2023yyx}.

Three new features of this process affect the complexity of the
reconstruction procedures:
\begin{itemize}
  \item the calculation assumes that the two Higgs bosons have equal
    mass, therefore an extra equivalence relation, besides momentum
    conservation, has to be imposed;
  \item the presence of {\it two} massive particles complicates the
    construction of a minimal ansatz, since the dependence on both
    their four-momenta cannot be removed by momentum conservation;
  \item the dependence on the mass of the quark in the loop occurs,
    not just as a Taylor series, but also mixed with kinematic poles.
\end{itemize}
We address these points in this section, with further details given in
Appendices \ref{sec:spinor_decompositions} and
\ref{sec:boxmanipulation}. This was required to arrive at a form of
the coefficients simple enough for them to be presented in this
article.

\subsection{Spinor variables subject to additional constraints}
In the first instance, the massive five-point process under
consideration can be embedded into a seven-point massless
process. Thus, Lorentz-covariant polynomials can be taken to belong to
the following polynomial ring,
\begin{equation}\label{eq:S7}
  S_7 = \mathbb{F} \big[ |1\rangle, [1|, \dots |7\rangle, [7| \big] \, .
\end{equation}
Legs $4,5$ and $6,7$ can be thought of as fictitious massless
scalar decay products of the two Higgs bosons.

To account for equivalence relations, such as momentum conservation
$\sum_i |i\rangle[i|$, we introduce a polynomial quotient ring. The
  quotient ring construction of ref.~\cite{DeLaurentis:2022otd} can be
  easily modified to account for the extra relation on the Higgs
  masses,
\begin{equation}\label{eq:R7}
  R_7 = S_7 \big / \big\langle \sum_{i=1}^{7} |i\rangle[i|,
    s_{45}-s_{67} \big\rangle \, .
\end{equation}

When reconstructing the integral coefficients, it is crucial to work
within the correct quotient ring.  For example, consider the box
coefficient $d_{4\x1\x23}$ (defined later in
section \ref{sec:method}). If we were to forget the $s_{45}-s_{67}$
constraint, the least common denominator (LCD) as obtained from the
generalized unitarity computation would involve the following
invariants,
\begin{equation}\label{eq:example-lcd-without-correct-constraint}
\langle13\rangle,\; \langle23\rangle,\;  [23],\; \langle1|(4+5)|1],\; \langle2|(4+5)|1],\; \langle3|(4+5)|1],\; [1|(4+5)|(6+7)|1]^2 \, ,
\end{equation}
as well as two more invariants that also involve $m^2$, the mass the quark
in the loop. However, once we impose the extra constraint
$s_{45}=s_{67}$, we are left only with 3 out of the 7 singularities of
eq.~\eqref{eq:example-lcd-without-correct-constraint},
\begin{equation}\label{eq:example-lcd-with-correct-constraint}
\langle23\rangle,\;  [23],\; [1|(4+5)|(6+7)|1]^2 \, .
\end{equation}
This is a drastic simplification. Furthermore, if we were to
analytically reconstruct the numerators of these four extra poles,
there is no guarantee that these would yield simple expressions, given
that the extra four poles effectively have residues proportional to
$(s_{45}-s_{67})$, meaning they could be arbitrarily complicated
rewritings of zero.

\subsubsection{Univariate interpolation with additional constraints}
\label{univariate-interpolation}
A key step in the reconstruction of multivariate rational functions is
the determination of their irreducible denominator factors. This can
be achieved by univariate Thiele interpolation \cite{Peraro:2016wsq}
on a generic univariate slice in the multivariate space
\cite{Abreu:2018zmy}. When spinor variables are involved, such a
univariate slice can be constructed from a BCFW shift
\cite{Britto:2005fq, Risager:2005vk} simultaneously applied to all
holomorphic or anti-holomorphic spinors \cite{Elvang:2008vz}, i.e.~an
\textit{all-line shift},
\begin{equation}\label{eq:holomorphic-shift}
  |i\rangle \rightarrow |i\rangle + t \, c_i |\eta\rangle, \quad [i| \rightarrow [i| \, ,
\end{equation}
for a given constant $|\eta\rangle$ and with generic $c_i$'s that
satisfy momentum conservation,
\begin{equation}
  \sum_i c_i |\eta\rangle = 0 \, .
\end{equation}
A single such slice suffices to determine the denominators of
functions of Mandelstam invariants and $\trfive={\rm Tr}\,
\{\slsh{p}_1\,\slsh{p}_2\,\slsh{p}_3\,\slsh{p}_4 \gamma_5\}$
\cite{Abreu:2021asb}, while the holomorphic plus anti-holomorphic pair
is required to obtain denominators of spinor-helicity functions
\cite{Abreu:2023bdp}, since we need the univariate slice to intersect
all codimension one varieties. The slice of
eq.~\eqref{eq:holomorphic-shift} does not intersect the purely
anti-holomorphic varieties associated to ideals generated by just
square brackets, $[ij]$.

Here we show that a single slice (instead of a pair) suffices, and
that additional constraints can also be imposed. To this end, we
construct a shift that involves all variables,
\begin{equation}\label{eq:generic-shift}
  |i\rangle \rightarrow |i\rangle + t \, x_i |\eta\rangle, \quad [i|
    \rightarrow [i| + t \, y_i [\eta| \, .
\end{equation}
Now momentum conservation takes the form,
\begin{equation}
  \sum_i \Big( |i\rangle [i| + t \big( x_i |\eta\rangle[i| + y_i |i\rangle [\eta| \big) + t^2 x_iy_i |\eta\rangle[\eta|  \Big) = 0 \, ,
\end{equation}
where the $t^0$ coefficient is automatically zero, as the starting
point is assumed to be in $R_7$.  Moreover, the extra constraint
$s_{45}-s_{67}$ becomes,
\begin{equation}
  \Big[ - \Big( \langle 45\rangle + t x_4 \langle \eta 5 \rangle + t x_5 \langle 4\eta \rangle \Big) \times \Big( \langle\rangle \leftrightarrow [] \Big) \Big] - \Big[ \{4,5\} \leftrightarrow \{6, 7\} \Big] = 0 \, ,
\end{equation}
which, once expanded, yields a quadratic polynomial in $t$, where,
once again, the $t^0$ coefficient is zero.

Collecting all coefficients in the $t$-polynomials, which in this case
means the $t$ and $t^2$ coefficients, we obtain a system of equations
in the variables $x_i, y_i$. A generic solution to this system ensures
the shifted line of eq.~\eqref{eq:generic-shift} lies entirely within
$R_7$ and crosses all relevant codimension-one varieties.  This allows
the determination of the LCD by matching irreducible denominator
factors in $t$ to a list of expected singularities.

\paragraph{Implementation}
We implement the univariate slice using \texttt{lips}
\cite{DeLaurentis:2023qhd, lips}, with
the numeric part relying on \texttt{NumPy} \cite{2020NumPy-Array}, and
the analytic part on \texttt{SymPy}
\cite{Meurer_SymPy_symbolic_computing_2017}. The package
\texttt{pyadic} \cite{pyadic} is used for
its implementation of the number field $\mathbb{F}$, taken to be
either finite fields ($\mathbb{F}_p$) or $p\kern0.2mm$-adic number,
($\mathbb{Q}_p$), and of the Thiele and Newton interpolation
algorithms \cite{Peraro:2016wsq}. The generic solution to the system
of equations in $\mathbb{F}\big[x_i, y_i \big]$ is obtained by a
generalization of the algorithm presented in ref.~\cite[Section
  3]{DeLaurentis:2022otd} for arbitrary polynomial (quotient) rings,
as implemented in \texttt{syngular}
\cite{syngular}
(\texttt{Ideal.point\_on\_variety}). The required lexicographic
Gr\"obner bases are obtained from \texttt{Singular} \cite{DGPS}.

\paragraph{Arbitrary quotient ring}
It appears clear that the same procedure could be applied when working
in an arbitrary quotient ring,
\begin{equation}
  \mathbb{F} \big[ \underline{X} \big] \big / \big\langle q_1(\underline{X}), \dots, q_m(\underline{X}) \big\rangle
\end{equation}
by performing a shift,
\begin{equation}
  \underline{X} \rightarrow \underline{X} + t \, \underline{Y} \, ,
\end{equation}
and substituting it into the ideal $\big\langle q_1, \dots, q_m
\big\rangle$. Then, the equations need to be expanded in $t$ and each
coefficient in the $t$-polynomials needs to be set to zero by choosing
appropriate solutions for $\underline{Y}$, such that the line lies in
the quotient ring for arbitrary values of $t$.

This procedure has its limits. Namely, if the polynomial quotient ring
is not a unique factorization domain, i.e.~if there exist irreducible
polynomials that generate non-prime ideals, then one has to be careful
with determining the LCD, as it is not unique. This is actually
relevant for the determination of the effective pentagon coefficients
discussed in section~\ref{sec:decomposition}

\subsection{Minimal spinor ansatz for an arbitrary number of massive scalars}

Minimal ans\"atze for the analytic reconstruction of an $n$-point
process with a single massive external scalar leg, such as single
Higgs production in association with jets \cite{Budge:2020oyl}, can be
constructed by considering an $(n-1)$-point process without momentum
conservation. This amounts to replacing every occurrence of the
massive four-momentum with a sum over all the massless ones. However,
this is no longer possible in the presence of multiple massive lines,
since removing a massive four-momentum causes the introduction of
another one.

Since the rings of eq.~\eqref{eq:S7} and eq.~\eqref{eq:R7}
over-parametrise the space of the $pp\rightarrow HHj$ coefficients,
to construct a minimal ansatz, we need to consider the covariant ring
without fictitious decays,
\begin{equation}\label{eq:massive-polynomial-ring}
  \bold{S}_5 = \mathbb{F} \big[ |1\rangle, [1|, |2\rangle, [2|, |3\rangle, [3|, \four, \five \big] \, ,
\end{equation}
where the bold numbers denote rank two spinors, $\four =
\four_{\alpha\dot\alpha}$ and $\five = \five_{\alpha\dot\alpha}$. The
relation between the seven massless legs of eq.~\eqref{eq:S7} and
three massless plus two massive ones of
eq.~\eqref{eq:massive-polynomial-ring} is,
\begin{equation}
  1 \rightarrow 1, \, 2 \rightarrow 2, \, 3 \rightarrow 3, \, \bold 4 \rightarrow 4+5, \, \bold 5 \rightarrow 6+7 \, .
\end{equation}

Because the external massive particles are scalars, the rank-two
spinors never appear decomposed as sums over two pairs of rank-one
spinors, $\bold{p}=\sum_I |\bold{p}^I\rangle[\bold{p}_I|$. In fact,
  this would be equivalent to the seven-point space of
  eqs.~\eqref{eq:S7} and \eqref{eq:R7}. It introduces 8 spinor
  components in lieu of 4.

The four, four-momentum conservation equations read\,
\begin{equation}\label{eq:mom-cons-ideal}
  J_{\text{mom.~cons.}} = \Big\langle  |1\rangle[1| + |2\rangle[2| + |3\rangle [3| + \four + \five \Big\rangle \,.
\end{equation}
We further impose the constraint that $m_{\four} = m_{\five}$. In terms
of the ring variables, the masses can be written as
$\tr(\four|\four)=2m^2_{\four}$, or $\text{det}(\four)=m^2_{\four}$,
and equivalently for $\five$. These are to be understood as
polynomials in the four components of the rank-two spinors.

The quotient ring is then,
\begin{equation}
  \bold{R}_5 = \bold{S}_5 / ( J_{\text{mom.~cons.}} + \left\langle \text{det}(\four) - \text{det}(\five) \right\rangle ) \,.
\end{equation}

Since amplitudes are Lorentz invariant, we now need to convert to an
invariant ring following the elimination algorithm of ref.~\cite[Section
  2]{DeLaurentis:2022otd}. However, the standard spinor brackets,
\begin{equation}\label{eq:massless_massless_variables}
  \langle ij \rangle \quad \text{and} \quad [ij] \, ,
\end{equation}
are no longer sufficient, since $i,j$ can only be in $\{1,2,3\}$. In
addition, we must consider contractions involving the rank-two spinors
$\four$ and $\five$, both with the rank-one spinors and among
themselves. The former case reads,
\begin{equation}\label{eq:massless_massive_variables}
  \langle i | \bold{k} |i] \quad \text{and} \quad \langle i | \bold{k} |j]\,,
\end{equation}
while the latter case
requires the introduction of traces, now considered as variables in
their own right (i.e.~no longer as polynomials in the components),
\begin{equation}\label{eq:massive_massive_variables}
  \tr(\bold{k}|\bold{k}) \quad \text{and} \quad \tr(\bold{k}|\bold{l}) \, .
\end{equation}
with $\bold{k}, \bold{l}$ in $\{\four, \five\}$.  Since we work with
2-component spinors, these are understood as traces of $2 \x 2$
matrices, explicitly
$\tr(\bold{k}|\bold{l})=\bold{k}_{\alpha\dot\alpha}\bold{l}^{\dot\alpha\alpha}$.
In terms of Mandelstam invariants and $\mh$ they read
$\tr(\four|\four) = 2 \, \mh^2$, $\tr(\four|\five) = s_{123} - 2 \,
\mh^2$.

The invariants of eq.~\eqref{eq:massless_massless_variables},
eq.~\eqref{eq:massless_massive_variables}, and
eq.~\eqref{eq:massive_massive_variables} altogether can be used to
define a Lorentz invariant polynomial ring analogous to
$\mathcal{S}_n$ of ref.~\cite[Section 2.2,
  Eq.~2.59]{DeLaurentis:2022otd},
\begin{equation}
  \mathcal{S}\kern-2.48mm\mathcal{S}_5 = \mathbb{F} \big[ \langle ij \rangle, [ij], \langle i | \bold{k} |i], \langle i | \bold{k} |j], \tr(\bold{k}|\bold{k}) , \tr(\bold{k}|\bold{l}) \big] \, .
\end{equation}
Once again these variables are subject to equivalence relations, thus
a polynomial quotient ring, $\mathcal{R}\kern-2.90mm\mathcal{R}_5$, is
needed. The equivalence relations now include Schouten identities,
besides momentum conservation and the equality between the two Higgs
masses.

We can obtain all these relations among Lorentz invariant spinor
brackets from those among the Lorentz covariant spinors. To do this,
we build an extended covariant plus invariant polynomial ring with the
variables from both $\bold{S}_5$ and
$\mathcal{S}\kern-2.48mm\mathcal{S}_5$. Following the notation of
ref.~\cite{DeLaurentis:2022otd} we have,
\begin{equation}
  \boldsymbol{\Sigma}_5 = \mathbb{F} \big[ |1\rangle, [1|, |2\rangle,
      [2|, |3\rangle, [3|, \four, \five , \langle ij \rangle, [ij],
          \langle i | \bold{k} |i], \langle i | \bold{k} |j],
      \tr(\bold{k}|\bold{k}) , \tr(\bold{k}|\bold{l}) \big] \, .
\end{equation}
In this extended ring, we
consider the ideal
\begin{align}
  \boldsymbol{\kappa}[J_{\text{mom.~cons.}} + \left\langle
    \text{det}(\four) - \text{det}(\five) \right\rangle] =& \; \Big\langle
  |1\rangle[1| + |2\rangle[2| + |3\rangle [3| + \four + \five, \nn
        \\
        & \qquad \text{det}(\four) - \text{det}(\five), \langle 12\rangle -
        (\lambda_{1,0}\lambda_{2,1}-\lambda_{2,0}\lambda_{1,1},\,
        \dots) \Big\rangle \, ,
\end{align}
where the ellipsis contains all other equations defining the invariant
spinor brackets as contractions of the components of the covariant
spinors. For clarity, we have expanded the definition of the $\langle
12\rangle$ bracket in terms of the components of the spinors
$|1\rangle = (\lambda_{1,0}, \lambda_{1,1})$ and $|2\rangle =
(\lambda_{2,0}, \lambda_{2,1})$. The minus sign arises from the
Levi-Civita tensor, which is the metric in spinor space.

The invariant quotient ring, $\mathcal{R}\kern-2.90mm\mathcal{R}_5$,
is obtained as
\begin{align}
  \mathcal{R}\kern-2.90mm\mathcal{R}_5 = \mathcal{S}\kern-2.48mm\mathcal{S}_5 / \big(\boldsymbol{\kappa}[J_{\text{mom.~cons.}} + \left\langle
    \text{det}(\four) - \text{det}(\five) \right\rangle] \cap \mathcal{S}\kern-2.48mm\mathcal{S}_5 \big) \, ,
\end{align}
where the intersection of the ideal $\boldsymbol{\kappa}$ of
$\boldsymbol{\Sigma}_5$ with the subring
$\mathcal{S}\kern-2.48mm\mathcal{S}_5$ amounts to eliminating the
spinor component variables. That is, within a suitably chosen
block-order, one picks the subset of the Gr\"obner basis generated
only by Lorentz invariant variables. This automatically generates all
equivalence relations among invariants.  They are: a) the invariant
equivalent of $J_{\text{mom.~cons.}}$, i.e.~the ideal generated by all
contractions of the covariant generator of
eq.~\eqref{eq:mom-cons-ideal}; b) $\tr(\four|\four)=\tr(\five|\five)$,
which comes from $\text{det}(\four) = \text{det}(\five)$; c) all
Schouten identities.  These are no longer the usual ones,
\begin{equation}
\langle ij \rangle\langle kl \rangle+\langle ik \rangle\langle lj \rangle+\langle il \rangle\langle jk \rangle = 0 \, ,
\end{equation}
since we have only three massless legs, instead we have ones of the
form,
\begin{equation}
  \spa{j}.{k} \spab{i}.{\bold{l}}.{k}-\spa{i}.{k} \spab{j}.{\bold{l}}.{k} + \spa{i}.{j} \spab{k}.{\bold{l}}.{k} = 0 \, .
\end{equation}

Constructing an ansatz then amounts to enumerating all independent
monomials of $\mathcal{R}\kern-2.90mm\mathcal{R}_5$ with a given mass
dimension and little-group weight. Mass dimension and polynomial
degree now are related by a weight vector assigning weight 1 to the
two-particle spinor brackets, and weight 2 to the three-particle
spinor brackets and traces. Polynomials are still homogeneous, given
this weight vector.

It is clear that this construction could be generalized to construct
ans\"atze for an arbitrary combination of $m$ massless legs plus $n$
massive scalar ones.

To conclude, we report the size of various ans\"atze in table
\ref{tab:ansatz-sizes}. The ansatz sizes for the process of interest
in this work is given in row 6, while row 7 shows the closely related
process where the two Higgs bosons are not constrained to have the
same mass. Comparing row 3 with row 6 demonstrates the importance of
using the appropriate ansatz construction for the present
calculation. In fact, while the ansatz of row 3, representing
polynomials in $\mathcal{R}_7$, would have worked, it clearly becomes
orders of magnitude more complex then the minimal one for this process
in row 6.

\begin{table}[t!]
    \centering
    \begin{tabular}{l|c|c|c|c|c|c}
         mass dimension: & 2 & 4 & 6 & 8 & 10 & 12 \\
        \hline
        1. $0 \rightarrow ggggg$ & 5 & 16 & 40 & 85 & 161 & 280 \\
        \hline
        2. $0 \rightarrow gggggg$ & 9 & 50 & 205 & 675 & 1886 & 4644 \\
        \hline
        3. $0 \rightarrow ggggggg$ & 14 & 120 & 735 & 3486 & 13566 & 45178 \\
        \hline
        4. $0 \rightarrow ggggH$ & 6 & 22 & 62 & 147 & 308 & 588 \\
        \hline
        5. $0 \rightarrow gggggH$ & 10 & 60 & 265 & 940 & 2826 & 7470 \\
        \hline
        6. $0 \rightarrow gggHH$ & 6 & 22 & 62 & 147 & 308 & 588 \\
        \hline
        7. $0 \rightarrow gggHH^*$ & 7 & 29 & 91 & 238 & 546 & 1134 \\
    \end{tabular}
    \caption{Ansatz dimensions at zero phase weights for various
      processes representative of external kinematic configurations. The
      first column corresponds to the number of independent Mandelstam
      invariants. Rows 1, 4, 6 and 7 involve a single independent
      $\trfive$, thus their ansatz sizes saturate the upper bound of
      ref.~\cite[Eq.~3.2]{DeLaurentis:2020qle}. Row 2 reproduces
      ref.~\cite[Table 1]{DeLaurentis:2019bjh} (implementations
      differ). Rows 4 and 6 are identical, but represent different
      ans\"atze. For instance, the former may have non-zero
      little-group weight associated with the fourth leg, but the
      latter cannot.}
    \label{tab:ansatz-sizes}
\end{table}

\paragraph{Implementation}
We implement the ansatz construction via the described adaptation of
the algorithm of ref.~\cite[Section 2.2]{DeLaurentis:2022otd} using
\texttt{lips} for the spinor algebra \cite{DeLaurentis:2023qhd, lips},
\texttt{Singular} for the Gr\"obner bases and variable elimination
\cite{DGPS}, and the \texttt{CP-SAT} solver from \texttt{OR-Tools}
\cite{ortools} to enumerate the ansatz monomials.

\paragraph{Massive vector bosons}

A similar construction, with appropriate modification, should be
suitable to build minimal ans\"atze for processes involving multiple
bosons, including vector ones. In a soon-to-appear publication
\cite{DeLaurentis:2024xxx}, the two-loop amplitudes for the process
$pp\rightarrow Wjj$ are reconstructed in spinor-helicity variables. To
do so, the ansatz for a single massive scalar (row 4 of Table
\ref{tab:ansatz-sizes}) is modified to allow the spinors of the now
non-fictitious decay products, e.g.~a charged lepton plus neutrino
pair, to appear with a degree bound of one (for more details see that
article). The two constructions could be combined with suitable degree
bounds on the leptonic decays of the vector bosons to construct
ans\"atze for processes involving e.g.~$WW$, $WZ$, $WH$, $ZZ$ or
$ZH$. The computation of two-loop amplitudes for such processes in
association with a jet is now within reach \cite{Abreu:2024flk}.

\subsection{Projective space in $\boldsymbol{m}$}\label{sec:prjectivem}

Besides the kinematic variables discussed so far, the mass of the
quark running in the loop also appears in the amplitude. The
polynomial ring is thus,
\begin{equation}
  \bold{S}_5[m] = \mathbb{F} \big[ |1\rangle, [1|, |2\rangle, [2|, |3\rangle, [3|, \four, \five, m \big] \, .
\end{equation}
The quotient ring construction is unchanged, since no equivalence
relation involves $m$, so we simply have the covariant quotient ring
$\bold{R}_5[m]$ and the invariant one
$\mathcal{R}\kern-2.90mm\mathcal{R}_5[m]$.

It is clear that $m$ has a special role in $\bold{R}_5[m]$ and
$\mathcal{R}\kern-2.90mm\mathcal{R}_5[m]$: it is the only variable not
subject to equivalence relations. It is then straightforward to
consider this space as being projective in $m$, since $m$ can be taken
to infinity independently. On the other hand, we remain in an affine
space for the spinors, since taking a spinor to infinity would require
another one also being either large (if additive in the equivalence
relation) or small (if multiplicative). That is, we include the point
$m\rightarrow\infty$, but not any point at infinity for the spinor
variables. In a projective space at the point at infinity the role of
poles and zeros is inverted, a numerator factor is a pole, while a
denominator factor is a zero.

The box coefficients in the $gggHH$ amplitude have a rich analytic
structure in $m$. Since performing four unitarity cuts can result in a
fifth propagator being evaluated on the cut, the kinematic and $m$
dependence mix in the denominator. The large and small $m$ limits
provide useful information on the general $m$ expression, but do not
directly translate to individual terms of a Taylor expansion as in the
case of the triangle coefficients. In Appendix
\ref{sec:boxmanipulation} we investigate how the loop quark mass
dependence affects various rewritings of a box coefficient, especially
in relation to the location of spurious singularities. In particular,
we make the interesting observation that there can be a spurious pole
at $m\rightarrow \infty$. This could provide an even more numerically
efficient way to write the box coefficients, if the size of the
expressions in that form can be contained. In fact, the mass $m$, when
interpreted as the top-quark mass, is large, but never truly
approaches infinity---unlike some kinematic variables that may need to
approach zero. The compact form of the box coefficients that we
present in section~\ref{sec:analytic} is instead based on an effective
pentagon decomposition, as discussed in section~\ref{sec:decomposition}.

\subsubsection{Effective pentagons as residues of mixed $\boldsymbol{m}$-kinematic poles}

There are two singularities (plus permutations) that mix the kinematic
dependence with the $m$ dependence. Their origin is explained in
section \ref{sec:pentbox}. For the purpose of the present discussion,
it is sufficient to anticipate the form of eq.~(\ref{Sdef2}) and
eq.~(\ref{SdefNonAdjacent}). For convenience, we repeat one here,
\begin{equation}
  |S^{1\x2\x3\x4}| = -s_{12}s_{23}\spaba1.\five.\four.3\spbab3.\four.\five.1+ m^2 \, \big(\trfive\big)^2 \, .
\end{equation}
An advantage of the decomposition in eq.~(\ref{pentagonreduction}) is
that the entire dependence on such mixed mass-kinematic poles is
captured by the coefficients $\mathcal{C}$, leaving the effective
pentagons $\hat e_{i\times j \times k \times l}$ and effective boxes
$\hat d_{i\times j \times k}$ free of any $m$ dependence in the
denominator.

The effective pentagons $\hat e_{i\times j \times k \times l}$ are
defined as residues of these mixed mass-kinematic singularities (up to
the numerator part of the $\mathcal{C}$'s). The residue of a simple
pole in $|S^{1\x2\x3\x4}|$ is defined in the quotient ring,
\begin{equation}\label{pentagon-q-ring}
  \bold{S}_5 / ( J_{\text{mom.~cons.}} + \left\langle m^2_{\four} -
  m^2_{\five} \right\rangle + \left\langle |S^{1\x2\x3\x4}|
  \right\rangle ) \, .
\end{equation}
This is not a unique factorization domain, because we can find an
irreducible polynomial, e.g.~$\trfive$, that generates a non-prime
ideal. For instance, in the quotient ring of
eq.~(\ref{pentagon-q-ring}), we have
\begin{equation}
  \left\langle \trfive \right\rangle \ni m^2 \trfive^2 = s_{12}s_{23}\spaba1.\five.\four.3\spbab3.\four.\five.1 \, .
\end{equation}
Since none of the factors $\langle 12\rangle$, $[12]$, $\langle
23\rangle$, $[23]$, $\spaba1.\five.\four.3$, and
$\spbab3.\four.\five.1$ belongs to $\left\langle \trfive
\right\rangle$ this proves that $\left\langle \trfive \right\rangle$ is not
prime (in the quotient ring of eq.~\eqref{pentagon-q-ring}).

This implies that the LCD of an effective pentagon coefficient is not
unique. In practice, we can shift its definition either additively by,
\begin{equation}\label{eff-pent-additive-redundancy}
  0 = -s_{12}s_{23}\spaba1.\five.\four.3\spbab3.\four.\five.1+ m^2 \,
  \big(\trfive\big)^2 \, ,
\end{equation}
or multiplicatively by,
\begin{equation}\label{eff-pent-multiplicative-redundancy}
  1 = \frac{m^2 \,
    \big(\trfive\big)^2}{s_{12}s_{23}\spaba1.\five.\four.3\spbab3.\four.\five.1} \,,
\end{equation}
without affecting its validity in relation to
eq.~(\ref{pentagonreduction}) and eq.~(\ref{Pentagondecomposition}).

We use the redundancy of eq.~(\ref{eff-pent-additive-redundancy}) to
show that the effective pentagons can be written without a double pole
in $\trfive$. This step was crucial to obtain expressions compact
enough to be presented in this article. The redundancy of
eq.~(\ref{eff-pent-multiplicative-redundancy}) could be used to
replace the spurious simple pole in $\trfive$ with a zero in
$\trfive$, but with extra spurious poles in
$s_{12},\,s_{23},\,\spaba1.\five.\four.3,\,\spbab3.\four.\five.1$. Such
re-definitions affect the form of the effective box coefficients, as
these are not defined as residues of the same pole.

\subsubsection{Interpolation on leading $\boldsymbol{p}\kern0.2mm$-adic digit}
When reconstructing functions that depend on $m$, it is generally
useful to consider the behaviour at $m = 0$ and $m \rightarrow
\infty$. For functions where the $m$ dependence is a Taylor series,
this allows the isolation of individual terms in the series. For functions
that have $m$ dependence mixed with kinematic variables in the LCD,
the small and large $m$ limits still yield useful information, but
cannot be directly used to obtain the general $m$ expression.

It is interesting to note that the univariate interpolation of section
\ref{univariate-interpolation}, traditionally used with finite fields,
can be equally well performed on a leading $p\kern0.2mm$-adic digit,
by treating it as if it was a $\mathbb{F}_p$ number
\cite{Chawdhry:2023yyx}. This allows one to isolate the $m=0$
contribution by setting $m\propto p$, and $m\rightarrow \infty$ by
setting $m \propto \frac{1}{p}$. If we were to work in $\mathbb{F}_p$
this would require interpolation on a bi-variate slice. Since the
ideal $\langle m \rangle$ is prime, the LCDs at $m = 0$ and $m
\rightarrow \infty$ are uniquely determined by the slicing procedure.

\section{Results for the process $0 \to q \qb g H H$}
\subsection{Process $0 \to q \qb g H^*$}
We first report on the contribution to Higgs boson pair production via the triple Higgs boson coupling,
due to the process,
\begin{equation}
  0 \to  q(p_{1}) + \qb(p_2)+g(p_3) + H^*(p_{4})\,,\;\;p_4=-p_1-p_2-p_3\,,\;\; p_4^2\neq \mh^2\,.
\end{equation}
The amplitude is given by,
\begin{equation}
  -i {\cal{A}}^{C_3}_{i_1 i_2}(1_q, 2_{\qb},3_g; 4_{H^*})=-\frac{g_s^3}{16 \pi^2} \frac{g_W}{2M_W} t^{C_3}_{i_1 i_2}
  A(1_q^{h_1}, 2_{\qb}^{h_2},3_g^{h_3}; 4_{H^*})\,.
\end{equation}
The $t$ matrices are the $SU(N)$ matrices in the fundamental representation normalized such that,
\begin{equation}    \label{normalization}
        \tr(t^a t^b)\;=\; \delta^{ab}\, ,
\end{equation}
and $N=3$. $i_1,i_2$ and $C_3$ are thus the $SU(3)$ indices of the quark, antiquark and gluon respectively,
The helicities are denoted by,
$h_1,h_2$ and $h_3$ for the outgoing quark, anti-quark and gluon respectively.
The result for the colour-stripped amplitude is,
\begin{equation}
  A(1_q^{-}, 2_{\qb}^{+}, 3_g^+; 4_{H^*})  =-4 \frac{\spa1.2 \spb2.3^2 m^2 }{(s_{123}-s_{12})^2} \Big[
       \Bzerobar(p_{12})
       +\Big[\frac{(s_{123}-s_{12})}{2s_{12}} -2 \frac{m^2}{s_{12}}\Big]\Czerobar(p_3,p_{12})
       -\frac{(s_{123}-s_{12})}{s_{12}}\Big]\,.
\label{eq:qqbgH}
\end{equation}
For conciseness we have introduced modified (dimensionless) forms of the scalar integrals,
\begin{eqnarray} \label{reduced_integrals}
\Czerobar(p_3,p_{12})&=&\left(2 p_3 \cdot p_{12}\right) C_0(p_3,p_{12};m)\nn \\
\Bzerobar(p_{12}) &=& B_0(p_{12};m)-B_0(p_{123};m)\, .
\end{eqnarray}
The amplitudes for the Higgs boson pair production cross section due to the triple Higgs boson coupling are simply related to the above result,
\begin{equation}
  {\cal{A}}^{C_3}_{i_1 i_2}(1_q, 2_{\qb},3_g; H^* \to 4_H, 5_H)  =
    \frac{3 g_W}{2 M_W} \, \kappa_\lambda \, \frac{\mh^2}{s_{123}-\mh^2 }\,
     {\cal{A}}^{C_3}_{i_1 i_2}(1_q, 2_{\qb},3_g; 4_{H^*}) \,.
\label{eq:HHHrelation}
\end{equation}
By combining eqs.~(\ref{eq:qqbgH}) and~(\ref{eq:HHHrelation}) one arrives at the result
for this contribution to the full amplitude, which is included in the expressions
in the following section.

\subsection{Process $0 \to q \qb g HH$}
The contribution to the full physical amplitude for the process,
\begin{equation}
 0 \to  q(p_{1}) + \qb(p_2)+g(p_3) + H(p_4) + H(p_5)\; ,
\end{equation}
is given by,
\begin{eqnarray} \label{qqbargHH}
  -i {\cal A}^{C_3}_{i_1 i_2}(1^{h_1}_{q},2^{h_2}_{\qb},3^{h_3}_{g};4_H,5_H) 
  = -\frac{g_s^3}{16 \pi^2} \frac{m^2}{v^2}}{\; t^{C_3}_{i_1 i_2}  \;  A(1^{h_1}_{q},2^{h_2}_{\qb},3^{h_3}_g;4_H,5_H)\,.
\end{eqnarray}
The colour-ordered sub-amplitudes in eq.~(\ref{qqbargHH}) can be expressed in terms of scalar integrals.
For the $0 \to q\qb g\Higgs\Higgs$ colour-stripped sub-amplitude we have,
\begin{eqnarray} \label{fermionreduction}
  A(1^{h_1}_{q},2^{h_2}_{\qb},3^{h_3}_{g};4_H,5_H) & = & \frac{\bar\mu^{4-n}}{r_\Gamma}\frac{1}{i \pi^{n/2}} \int {\rm d}^n \ell
 \, \frac{{\rm Num}(\ell)}{\prod_i d_i(\ell)} \nn \\
&=& \sum_{i,j,k} {d}_{i\x j\x k}(1^{h_1},2^{h_2},3^{h_3}) \, D_0(p_i, p_j, p_k ;m)  \nn \\
&+& \sum_{i,j} {c}_{i\x j}(1^{h_1},2^{h_2},3^{h_3}) \,  C_0(p_i,p_j ;m)   \nn \\
&+& \sum_{i} {b}_{i}(1^{h_1},2^{h_2},3^{h_3}) \, B_0(p_i;m) + r(1^{h_1},2^{h_2},3^{h_3})\, .
\end{eqnarray}
The scalar bubble ($B_0$), triangle ($C_0$), box ($D_0$) integrals, and the
constant $r_\Gamma$, are defined in Appendix~\ref{Integrals}.
$r(1^{h_1},2^{h_2},3^{h_3})$ is the rational contribution to the amplitude.
We can further decompose the box and triangle coefficients according to the power of the quark mass running in the loop,
\begin{eqnarray}
  d_{i\x j\x k} &=&  d^{(0)}_{i\x j\x k} + m^2 \, d^{(2)}_{i\x j\x k} + m^4 \, d^{(4)}_{i\x j\x k} \, ,\nn \\
  c_{i\x j} &=&  c^{(0)}_{i\x j} + m^2 \, c^{(2)}_{i\x j}\,.
\end{eqnarray}
We present results for the helicity choices $h_1=-1,h_2=+1,h_3=+1$. The other helicity choices are obtained as follows,
\beqn
A(1^+,2^-,3^+)&=&A(2^-,1^+,3^+) \nn \\
A(1^+,2^-,3^-)&=&A(1^-,2^+,3^+)\big|_{\langle \rangle \leftrightarrow [ ]} \, .
\eeqn

\subsubsection{Boxes}
There are two independent box coefficients. The first one reads,
\begin{eqnarray}
  d^{(0)}_{12\x4\x3}(1^-,2^+,3^+) &=& \mh^2\*(s_{3\four}\*s_{3\five}-\mh^4)\*\frac{(\spab3.\four.2^2+\spab3.\five.2^2)}{\spb1.2\spaba3.\four.\five.3^2} \,,\\
d^{(2)}_{12\x4\x3}(1^-,2^+,3^+) &=& 8\frac{\spab3.\four.2\spab3.\five.2 \*(s_{3\four}\*s_{3\five}-\mh^4)}{\spb1.2\spaba3.\four.\five.3^2} + \bar{d}^{(2)}(1^-,2^+,3^+)\,,\\
\text{with} \quad \bar{d}^{(2)}(1^-,2^+,3^+) &=&
4\frac{\spab1.3.2 (\spabab1.\four.(1+2).\five.1-\spabab1.\five.(1+2).\four.1)}{s_{12} \spaba3.\four.\five.3} \,, \nn \\
 &-& 2\frac{(\spabab1.\four.(1+2).\five.2-\spabab1.\five.(1+2).\four.2)(s_{\four\five}-2s_{23}-2\mh^2)}{s_{12} \spaba3.\four.\five.3}  \,,\\
d^{(4)}_{12\x4\x3}(1^-,2^+,3^+) &=& -8\frac{\spa1.3\spab1.(\four-\five).3}{\spa1.2\spaba3.\four.\five.3}-8\frac{\spb2.3\spab3.(\four-\five).2}{\spb1.2\spaba3.\four.\five.3}\, .
\end{eqnarray}
The second one reads,
\begin{eqnarray}
  d^{(0)}_{12\x3\x4}(1^-,2^+,3^+) &=&-(s_{3\four}\*s_{\four\five}-\mh^2\*s_{12})\*\frac{(\spaba1.\five.\four.3^2+ \spa1.3^2\*\mh^4)}{\spa1.2\*\spaba3.\four.\five.3^2}\,,\\
d^{(2)}_{12\x3\x4}(1^-,2^+,3^+) &=& -8\frac{\spa1.3\spaba1.\five.\four.3(s_{3\four}\*s_{\four\five}-\mh^2\*s_{12})}{\spa1.2\*\spaba3.\four.\five.3^2} + \bar{d}^{(2)}(1^-,2^+,3^+)\,,\\
d^{(4)}_{12\x3\x4}(1^-,2^+,3^+) &=& d^{(4)}_{12\x4\x3}(1^-,2^+,3^+)\,.
\end{eqnarray}
The third box coefficient is not independent,
\begin{eqnarray}
  d_{3\x12\x4}(1^-,2^+,3^+) &=& d_{12\x3\x4}(1^-,2^+,3^+)\big|_{\four\leftrightarrow\five}\,.
\end{eqnarray}

\subsubsection{Triangles}
In the limit $m \to 0$ the box integrals, and a subset of the triangle
integrals, develop poles in the limit $\epsilon \to 0$.
Since the result for the amplitude must be finite, this yields
constraints on the integral coefficients.
The order $m^0$ triangle coefficients are determined by these IR relations,
\begin{eqnarray}
    \frac{c^{(0)}_{3\x4}}{s_{3\four}-\mh^2} &=&
        -\frac{d^{(0)}_{12\x3\x4}}{s_{3\four}\,s_{\four\five}-\mh^2\,s_{12}}
        -\frac{d^{(0)}_{12\x4\x3}}{s_{3\four}\,s_{3\five}-\mh^4}\,,\\
   \frac{c^{(0)}_{3\x124}}{s_{3\five}-\mh^2} &=&
              -\frac{d^{(0)}_{3\x12\x4}}{s_{3\five}\,s_{\four\five}-\mh^2\,s_{12}}
              -\frac{d^{(0)}_{12\x4\x3}}{s_{3\four}\,s_{3\five}-\mh^4}\,, \\
    \frac{\bar{c}^{(0)}_{3\x12}}{s_{\four\five}-s_{12}}&=&
            -\frac{d^{(0)}_{3\x12\x4}}{s_{3\five}\, s_{\four\five}-\mh^2\,s_{12}}
            -\frac{d^{(0)}_{12\x3\x4}}{s_{3\four}\,s_{\four\five}-\mh^2\,s_{12}}\,,
\end{eqnarray}
where this last relation represents the triangle coefficient originating
from diagrams that do not involve the triple Higgs coupling.  The full result for this
coefficient is,
\begin{equation}
c^{(0)}_{3\x12}=\bar{c}^{(0)}_{3\x12} + 6 \kappa_\lambda \frac{\spb2.3^2}{\spb1.2}\frac{\mh^2}{(s_{\four\five}-\mh^2)}\,.
\end{equation}
The order $m^2$ pieces for the triangles $c_{3\x12},c_{3\x4}$ and
$c_{3\x124}$ with one light-like external line are,
\begin{eqnarray}
c^{(2)}_{3\x12}(1^-,2^+,3^+)&=&\frac{8\spa1.3^2 \*(s_{\four\five}-s_{12})\*(s_{\four\five}-2\mh^2)}{\spa1.2\,\*\spaba3.\four.\five.3^2}
+\frac{8\spb2.3^2}{\spb2.1\,\* (s_{\four\five}-s_{12})}\Big[1+\kappa_\lambda \frac{3 \mh^2}{(s_{\four\five}-\mh^2)}\Big]\,, \qquad \\
c^{(2)}_{3\x4}(1^-,2^+,3^+)&=& 8\frac{\spab3.\four.3}{\spaba3.\four.\five.3^2}
\Big\{ \frac{\spa1.3\spaba1.\five.\four.3}{\spa1.2} - \frac{\spab3.\four.2\spab3.\five.2}{\spb1.2} \Big\}\,, \\
c^{(2)}_{3\x124}&=& c^{(2)}_{3\x4}(1^-,2^+,3^+)\big|_{\four\leftrightarrow\five}\,.
\end{eqnarray}
The triangle coefficients for the triangles without a light-like external line are,
\begin{eqnarray}
  c_{4\x123}(1^-,2^+,3^+) &=& (s_{\four\five}-2\*\mh^2+8\*m^2) \nn \\
  &\times&\Big\{\frac{\spaba1.(2+3).{(\four-\five)}.3 \* \big(\spaba1.\four.\five.3-\spaba1.\five.\four.3\big)}{\spa1.2\, \spaba3.\four.\five.3^2}
  -2 \frac{\spaba1.\four.\five.1}{\spa1.2\, \spaba3.\four.\five.3}\Big\}\,,\\
c_{4\x12}(1^-,2^+,3^+) &=& (s_{\four\five}-2\*\mh^2+8\*m^2)
  \Big\{\frac{\spa1.3\spab3.\five.2\Delta_{12|\four|3\five}}{s_{12}\spaba3.\four.\five.3^2}
  -\frac{\spab1.\four.2 (s_{3\five}+s_{12}-\mh^2)}{s_{12}\spaba3.\four.\five.3}\Big\} \nn \\
  & +&\frac{\spb3.2\*\Big\{\spab2.\four.2 \*\big(\spaba1.(1+2).\four.3-\spaba1.\four.(1+2).3\big)\Big\}}{s_{12}\*\spaba3.\five.\four.3} \nn \\
  & -&\frac{\spb3.2\*\Big\{\spab1.\four.2 \*\big(\spaba2.(1+2).\four.3-\spaba2.\four.(1+2).3\big)\Big\}}{s_{12}\*\spaba3.\five.\four.3} \nn \\
     & +& \frac{\big(\spaba1.(1+2).\four.3-\spaba1.\four.(1+2).3\big)\*\spab1.\four.3}{\spa1.2\*\spaba3.\five.\four.3} \,,\\
  c_{12\x34}(1^-,2^+,3^+) &=& c_{4\x12}\big|_{\four\leftrightarrow\five}\,,
\end{eqnarray}
and the K\"{a}ll{\'e}n function is,
\begin{equation}\label{eq:delta12-4-35}
\Delta_{12|\four|3\five}=(s_{12}+\mh^2-s_{3\five})^2-4\*s_{12}\*\mh^2\, .
\end{equation}

\subsubsection{Bubbles and rational pieces}
The bubble coefficients and the rational terms are given by,
\begin{equation}
  b_{123}(1^-,2^+,3^+) = 4 \frac{\spa1.2\*\spb2.3^2}{(s_{\four\five}-s_{12})^2}\Big[1+\kappa_\lambda \frac{3 \mh^2}{(s_{\four\five}-\mh^2)}\Big]\, ,\;\;\;
  b_{12}(1^-,2^+,3^+) = -b_{123}(1^-,2^+,3^+)\,,
\end{equation}
\begin{equation}
  r(1^-,2^+,3^+) = \frac{4\*\spb2.3^2}{\spb2.1\*(s_{\four\five}-s_{12})} \Big[1+\kappa_\lambda \frac{3 \mh^2}{(s_{\four\five}-\mh^2)}\Big]\, .
\end{equation}
This concludes the discussion of the quark-antiquark-gluon contribution to the Higgs boson pair amplitudes.
\section{Results for the process $0\to gggH^*(\to HH)$}
We first present results for the process involving a single Higgs boson~\cite{Ellis:1987xu,Baur:1989cm}
which contribute to the Higgs boson pair production via the Higgs boson self-coupling,
\begin{equation}
i {\cal A}^{C_1 C_2 C_3}(gggH^*)  =  \big(i \sqrt{2} f^{C_1 C_2 C_3}\big) \frac{g_s^3}{16 \pi^2}\, \frac{g_W}{M_W} \, s_{123}^2 \, A(1_g^{h_1}, 2_g^{h_2}, 3_g^{h_3}, 4_{H^*})\,,
\end{equation}
where $p_4=-p_1-p_2-p_3$ and $s_{12}=(p_1+p_2)^2,\,s_{13}=(p_1+p_3)^2,\, s_{23}=(p_2+p_3)^2,\, s_{123}=(p_1+p_2+p_3)^2$.
With these definitions we have, 
\begin{eqnarray}
A(1_g^{+}, 2_g^{+}, 3_g^{+}, 4_{H^*}) &=& \frac{1}{\spa1.2 \spa 2.3 \spa3.1} A_4(p_1,p_2,p_3) \,,\\
A(1_g^{+}, 2_g^{+}, 3_g^{-}, 4_{H^*}) &=& \frac{\spb1.2}{\spa1.2^2 \spb 1.3 \spb2.3} A_2(p_1,p_2,p_3)\,. 
\end{eqnarray}
The remaining amplitudes can be obtained by symmetry operations,
\begin{eqnarray}
A(1_g^{+}, 2_g^{-}, 3_g^{+}, 4_{H^*}) &=& - A(1_g^{+}, 3_g^{+}, 2_g^{-}, 4_{H^*})\,,\\ 
A(1_g^{-}, 2_g^{+}, 3_g^{+}, 4_{H^*}) &=& - A(3_g^{+}, 2_g^{+}, 1_g^{-}, 4_{H^*})\,,\\ 
A(1_g^{-h_1}, 2_g^{-h_2}, 3_g^{-h_3}, 4_{H^*}) &=& 
 -\left[ A(1_g^{h_1}, 2_g^{h_2}, 3_g^{h_3}, 4_{H^*}) \right]_{\spa{}.{} \leftrightarrow \spb{}.{}} \,.
\end{eqnarray}
The two helicity amplitudes $A_2,A_4$ are given by~\cite{Ellis:1987xu},
\begin{eqnarray}
&&A_4(p_1,p_2,p_3)=\frac{m^2}{s_{123}} \Big[
       -2-\big(\frac{m^2}{s_{123}}-\frac{1}{4}\big)\Big\{\Dzerobar(p_1,p_2,p_3)+\Dzerobar(p_1,p_3,p_2)+\Dzerobar(p_2,p_1,p_3) \nn \\
       &+&2\*\Czerobar(p_1,p_{23})+2\*\Czerobar(p_2,p_{13})+2\*\Czerobar(p_3,p_{12})\Big\}\Big]\,, \\
&&A_2(p_1,p_2,p_3)=\frac{m^2}{s_{123}^2}\*\Big[
  \frac{s_{12}\*(s_{23}-s_{12})}{s_{12}+s_{23}}+\frac{s_{12}\*(s_{13}-s_{12})}{s_{12}+s_{13}} \nn \\
  &-&2\*s_{13}\*s_{23}\*\frac{(2\*s_{12}+s_{23})}{(s_{12}+s_{23})^2}\*\Bzerobar(p_{13})
  -2\*s_{13}\*s_{23}\*\frac{(2\*s_{12}+s_{13})}{(s_{12}+s_{13})^2}\*\Bzerobar(p_{23}) \nn\\
   &+&(m^2-\frac{s_{12}}{4})
  \*\Big\{ 2\*\Czerobar(p_2,p_{13})+2\*\Czerobar(p_1,p_{23})-2\*\Czerobar(p_3,p_{12})-\Dzerobar(p_1,p_2,p_3)-\Dzerobar(p_2,p_1,p_3)\Big\} \nn\\
  &-&2\*s_{12}^2\*\Big[\Big(\frac{2\* m^2}{(s_{12}+s_{23})^2}-\frac{1}{2 (s_{12}+s_{23})}\Big)\*\Czerobar(p_2,p_{13})
                 +\Big(\frac{2\* m^2}{(s_{12}+s_{13})^2}-\frac{1}{2 (s_{12}+s_{13})}\Big)\*\Czerobar(p_1,p_{23}\Big)\Big] \nn \\
    &+&\frac{s_{23}\*s_{13}}{s_{12}}\*\big(\Czerobar(p_2,p_{13})+\Czerobar(p_1,p_{23})-\Czerobar(p_1,p_3)-\Czerobar(p_2,p_3)\big)\nn \\
    &-&\frac{1}{4} \*\big(s_{12}-12\*m^2-\frac{4 s_{23}\*s_{13}}{s_{12}}\big)\*\Dzerobar(p_1,p_3,p_2)\Big]\,.
\end{eqnarray}
where the reduced scalar integrals $\Czerobar$ and $\Bzerobar$ are defined in eq.~(\ref{reduced_integrals}) and
\begin{equation}
\Dzerobar(p_1,p_2,p_3) = \left(4 p_1 \cdot p_2 \; p_2 \cdot p_3\right) D_0(p_1,p_2,p_3;m) \,. 
\end{equation}
We note that $A_2$ is symmetric under the exchange $p_1 \leftrightarrow p_2$, whereas $A_4$ is totally symmetric.

The amplitudes for the Higgs pair production cross section due to the triple Higgs boson coupling are simply related to the above result,
\begin{equation}
  {\cal A}^{C_1 C_2 C_3}(gggHH)  =  \frac{3 g_W}{2 M_W} \, \kappa_\lambda \, \frac{\mh^2}{s_{123}-\mh^2 }\, {\cal A}^{C_1 C_2 C_3}(gggH^*) \,.
\end{equation}
Putting this together we arrive at the final form,
\begin{eqnarray}
  i {\cal A}^{C_1 C_2 C_3}(gggHH)  &=&  \frac{3 g_W}{2 M_W} \, \kappa_\lambda \, \frac{\mh^2}{s_{123}-\mh^2 }\,
  \big(i \sqrt{2} f^{C_1 C_2 C_3}\big) \frac{g_s^3}{16 \pi^2}\, \frac{g_W}{M_W} \, s_{123}^2
   \, A(1_g^{h_1}, 2_g^{h_2}, 3_g^{h_3}, (4+5)_{H^*}) \nn \\
   &=& \frac{g_s^3}{4 \pi^2} \, \frac{m^2}{v^2} \, \big[\tr\,(t^{C_1}t^{C_2}t^{C_3})-\tr\,(t^{C_3}t^{C_2}t^{C_1})\big]
    H_\kappa(1^{h_1},2^{h_2},3^{h_3};\Higgs,\Higgs)\,,
\label{eq:gggHHtriple}  
\end{eqnarray}
where we have extracted the overall Yukawa coupling factor from $A$ for later convenience and defined the
auxiliary amplitude,
\begin{equation}
H_\kappa(1^{h_1},2^{h_2},3^{h_3};\Higgs,\Higgs) = 
 \frac{3}{2} \, \kappa_\lambda \, \frac{\mh^2}{s_{123}-\mh^2} \, s_{123}^2
 \, \frac{A(1_g^{h_1}, 2_g^{h_2}, 3_g^{h_3}, (4+5)_{H^*})}{m^2}\,.
\end{equation}

\section{Calculation methods for the process $0\to gggHH$}
\label{sec:method}

In this section we introduce calculation details for the part of the process $0\to gggHH$ which does not involve the triple Higgs coupling,
\begin{equation}\label{eq:gggHH-process-kinematics-definition}
  0 \to g(p_1) + g(p_2) + g(p_3) + \Higgs(p_4) +\Higgs(p_5)\,.
\end{equation}
Both Higgs bosons are radiated off the quark line, with $p_1^2=p_2^2=p_3^2=0$ and $p_4^2=p_5^2=\mh^2$.
The analytic results will be presented in section~\ref{sec:analytic}.
\subsection{Definition of colour amplitudes}
The amplitude for the production of a pair of Higgs bosons and $3$ gluons can be expressed as
colour-stripped sub-amplitudes as follows,
\begin{eqnarray}
        \label{exp}
        i {\cal H}_n^{ggg}(\{p_i,h_i\})\,&=& \frac{g_s^3}{4 \pi^2} \frac{m^2}{v^2}\; \big[\tr\,(t^{C_1}t^{C_2}t^{C_3})-\tr\,(t^{C_3}t^{C_2}t^{C_1})\big]
        H(1^{h_1},2^{h_2},3^{h_3};\Higgs,\Higgs)\, .
\label{eq:gggHHradiated}  
\end{eqnarray}
$m$ is the mass of the quark circulating in the loop, and $v$ is the vacuum expectation value. 
Squaring the amplitude and summing over colours we have,
\begin{eqnarray}
      \sum_{\text{colours}}  \left| {\cal H}_n^{ggg}(\{p_i,h_i\}) \right|^2\,
      &=& 2 V N \left( \frac{g_s^3}{4 \pi^2} \frac{m^2}{v^2} \right)^2
        \left| H (1^{h_1},2^{h_2},3^{h_3};\Higgs,\Higgs) \right|^2 \, ,
\end{eqnarray}
where $V=N^2-1$.
From eqs.~\eqref{eq:gggHHtriple} and \eqref{eq:gggHHradiated}
it is clear that we can account for all diagrams, including the triple Higgs boson
interaction, by simple modification,
\begin{eqnarray}
      \sum_{\text{colours}}  \left| {\cal H}_n^{ggg}(\{p_i,h_i\}) \right|^2\,
      &=& 2 V N \left( \frac{g_s^3}{4 \pi^2} \frac{m^2}{v^2} \right)^2
        \left| H(1^{h_1},2^{h_2},3^{h_3};\Higgs,\Higgs) + H_\kappa(1^{h_1},2^{h_2},3^{h_3};\Higgs,\Higgs) \right|^2 \,.
	\nn \\
\end{eqnarray}

\subsection{Decomposition into scalar integrals}
\label{Decomposition}
The colour-ordered sub-amplitudes can be expressed in terms of scalar integrals.  For the $0 \to ggg\Higgs\Higgs$
sub-amplitude we have,
\begin{eqnarray} \label{fermionreduction2}
  H(1^{h_1},2^{h_2},3^{h_3}; 4_{\Higgs} 5_{\Higgs}) & = & \frac{\bar\mu^{4-n}}{r_\Gamma}\frac{1}{i \pi^{n/2}} \int {\rm d}^n \ell
 \, \frac{{\rm Num}(\ell)}{\prod_i d_i(\ell)} \nn \\
&=& \sum_{i,j,k} {d}_{i\x j\x k}(1^{h_1},2^{h_2},3^{h_3}) \, D_0(p_i, p_j, p_k ;m)  \nn \\
&+& \sum_{i,j} {c}_{i\x j}(1^{h_1},2^{h_2},3^{h_3}) \,  C_0(p_i,p_j ;m)   \nn \\
&+& \sum_{i} {b}_{i}(1^{h_1},2^{h_2},3^{h_3}) \, B_0(p_i;m) + r(1^{h_1},2^{h_2},3^{h_3})\, .
\end{eqnarray}
The scalar bubble ($B_0$), triangle ($C_0$), box ($D_0$) integrals, and the
constant $r_\Gamma$, are defined in Appendix~\ref{Integrals}.
$\bar{\mu}$ is an arbitrary mass scale, and $r$ are the rational terms.
All scalar integrals are well known and readily evaluated
using existing libraries~\cite{Ellis:2007qk,vanHameren:2010cp,Carrazza:2016gav}.

In order to obtain concise analytic expressions, we found that it was expedient
to re-express the box coefficients in terms of scalar pentagon integrals and a remainder. 
In order to perform this separation we are forced to introduce a denominator factor
of $\trfive$, which is given by,
\begin{equation}
\trfive={\rm Tr}\, \{\slsh{p}_1\,\slsh{p}_2\,\slsh{p}_3\,\slsh{p}_4 \gamma_5\} \,.
\label{eq:tr5def}
\end{equation}
In infrared configurations involving $p_1$, $p_2$ and $p_3$ this factor vanishes, giving rise to
potential numerical issues in these limits.  However we have been able to eliminate all factors
except for a single pole in $\trfive$, which mitigates these issues to a large extent.

\subsection{Basis integrals}
In section \ref{sec:analytic} we will present results for a minimal set of integral coefficients.
The remaining coefficients can be simply related to these
by permutation of momentum labels.  Here we summarize the basic set and the permutations required
to generate all coefficients.

There are 2 independent bubbles:
\begin{itemize}
\item $B_0(p_{12};m)$ ($\times$ 3 perms)\,;
\item $B_0(p_{123};m)$\,.
\end{itemize}
There are 5 independent triangles:
\begin{itemize}
\item $C_0(p_{1},p_{2};m)$ ($\times$ 3 perms)\,;
\item $C_0(p_{1},p_{4};m)$ ($\times$ 3 perms)\,;
with $C_0(p_{1},p_{234};m)$ obtained by $4 \leftrightarrow 5$ (= 6 perms)\,;
\item $C_0(p_{3},p_{12};m)$ ($\times$ 3 perms)\,;
\item $C_0(p_{4},p_{12};m)$ ($\times$ 3 perms)\,;
with $C_0(p_{12},p_{34};m)$ obtained by $4 \leftrightarrow 5$ (= 6 perms)\,;
\item $C_0(p_{4},p_{123};m)$\,.
\end{itemize}
There are 5 independent boxes:
\begin{itemize}
\item $D_0(p_{1},p_{2},p_{3};m)$ ($\times$ 3 perms)\,;
\item $D_0(p_{1},p_{2},p_{4};m)$ ($\times$ 6 perms)\,;
with $D_0(p_{34},p_{1},p_{2};m)$ obtained by $4 \leftrightarrow 5$ (= 12 perms)\,;
\item $D_0(p_{1},p_{4},p_{23};m)$ ($\times$ 3 perms)\,;
\item $D_0(p_{1},p_{4},p_{2};m)$ ($\times$ 3 perms) \,;
with $D_0(p_{2},p_{34},p_{1};m)$ obtained by $4 \leftrightarrow 5$ (= 6 perms)\,;
\item $D_0(p_{4},p_{1},p_{23};m)$ ($\times$ 3 perms)\,;
with $D_0(p_{1},p_{23},p_{4};m)$ obtained by $4 \leftrightarrow 5$ (= 6 perms)\,.
\end{itemize}
Additional permutations correspond to either the three cyclic choices of
$(1,2,3)$, or to all six permutations.
Some of these coefficients vanish for particular helicity choices.

Furthermore we can limit ourselves to the calculation of coefficients
where no more than one gluon has positive helicity:
\begin{equation}
c(1^-, 2^-, 3^-; 4,5),\;\;c(1^-, 2^-, 3^+; 4,5),\;\;c(1^-, 2^+, 3^-; 4,5)\,~\mbox{and}~\,c(1^+, 2^-, 3^-; 4,5)\,,
\end{equation}
where $c$ represents any of the coefficients ${d}_{i\x j\x k}$, ${c}_{i\x j}$ or ${b}_{i}$.
This is because parity relates coefficients with opposite helicities,
\begin{equation}
c(1^{-h_1}, 2^{-h_2}, 3^{-h_3}; 4,5)
 = \left[ c(1^{h_1}, 2^{h_2}, 3^{h_3}; 4,5) \right]^* \,.
\end{equation}

\subsection{Strategy for integral coefficients}

\subsubsection{Bubbles and rational terms}
\label{sec:bubbles}

The bubble coefficient $b_{12}$ and rational term $R$ are computed by
a direct calculation using Passarino-Veltman reduction.  A generic tensor
is constructed with free indices $\mu_1$, $\mu_2$, $\mu_3$ corresponding to
the currents for each of the gluons, eliminating terms that vanish due to
gauge invariance, ($p_1^{\mu_1} = p_2^{\mu_2} = p_3^{\mu_3} = 0$) and
employing a cyclic choice of gauge,
($p_2^{\mu_1} = p_3^{\mu_2} = p_1^{\mu_3} = 0$).  The final result for the
tensor is then simply contracted with the appropriate polarization vectors.

The coefficients of the bubbles $b_{23}$ and $b_{13}$ are obtained by cyclic
permutation of $\{p_1,p_2,p_3\}$:
\begin{eqnarray}
  b_{23}(1^{h_1},2^{h_2},3^{h_3})&=&b_{12}(2^{h_2},3^{h_3},1^{h_1}) \, ,\\
  b_{13}(1^{h_1},2^{h_2},3^{h_3})&=&b_{12}(3^{h_3},1^{h_1},2^{h_2}) \,.
\end{eqnarray}
The remaining bubble coefficient is then determined by the ultra-violet finiteness of the amplitude,
\begin{equation}
b_{123}(1^{h_1},2^{h_2},3^{h_3})=
 -b_{12}(1^{h_1},2^{h_2},3^{h_3})-b_{23}(1^{h_1},2^{h_2},3^{h_3})- b_{13}(1^{h_1},2^{h_2},3^{h_3}) \,.
\end{equation}

\subsubsection{Triangles}

The triangle coefficient $c_{1\x2}$ and the $m^2$ contributions to $c_{3\x12}$
and $c_{1\x4}$ are calculated in the same fashion as
the bubble and rational contributions.
The triangle coefficients $c_{4\x12}$ and $c_{4\x123}$ are obtained by
the unitarity methods of Forde~\cite{Forde:2007mi} and subsequently simplified.
      
The $m^0$ contribution to $c_{3\x12}$ and $c_{1\x4}$ coefficients are obtained
through infrared relations.
We perform a decomposition of the triangle coefficients,
\begin{equation}
c_{A\x B} = c_{A\x B}^{(0)} + m^2 c_{A\x B}^{(2)} \,,
\end{equation} 
such that the first term is the result obtained when setting the mass of the circulating
fermion to zero, except in the Yukawa coupling to the Higgs bosons. 
We can again exploit the fact that the amplitude must be
infra-red finite in this limit to constrain the coefficients of 
box and triangle integrals that develop poles as $\epsilon \to 0$.  Specifically we
find that $c_{1 \x 4}^{(0)}$ is given by a combination of box coefficients,
\begin{eqnarray} \label{IR1}
\frac{c_{1 \x 4}^{(0)}}{(s_{1\four}-\mh^2)} &=&
   \frac{2 d_{3\x 14\x 2}^{(0)}}{(s_{2\five} \,s_{3\five}-s_{1\four} \, \mh^2)}
 - \frac{2 d_{1\x 4\x 3}^{(0)}} {(s_{3\four} \,s_{1\four}-s_{2\five} \, \mh^2)}
 - \frac{2 d_{2\x 4\x 1}^{(0)}} {(s_{1\four} \,s_{2\four}-s_{3\five} \, \mh^2)}
 - \frac{d_{23\x 4\x 1}^{(0)}}  {(s_{1\five} \,s_{1\four}-\mh^4)} \nn \\ 
 &-& \frac{d_{23\x 1\x 4}^{(0)}}{(s_{1\four} \, s_{\four\five}-\mh^2 \, s_{23})} 
 + \frac{d_{14\x 2\x 3}^{(0)}}{s_{23} \, s_{3\five}} 
 + \frac{d_{14\x 3\x 2}^{(0)}}{s_{23} \, s_{2\five}} 
 - 2\, \frac{d_{4\x 1\x 2}^{(0)}}{s_{12} \, s_{1\four}} 
 - 2\, \frac{d_{3\x 1\x 4}^{(0)}}{s_{13} \, s_{1\four}} \,.
\end{eqnarray}
A second relation determines $c_{3 \x 12}^{(0)}$ in terms of box coefficients
and the result for $c_{1 \x 2}^{(0)}$,
\begin{eqnarray} \label{IR2}
\frac{c_{3 \x 12}^{(0)}}{s_{13}+s_{23}} &=&
   \frac{c_{1\x 2}^{(0)}}{s_{12}} 
 - \frac{d_{3\x 12\x 4}^{(0)}}{s_{3\five} \, s_{\four\five}-\mh^2 \, s_{12}} 
 - \frac{d_{12\x 3\x 4}^{(0)}}{s_{3\four} \, s_{\four\five}-\mh^2 \, s_{12}} \nn \\
 &+&\frac{d_{1\x 2\x 4}^{(0)}}{s_{12} \, s_{2\four}} 
 + \frac{d_{4\x 1\x 2}^{(0)}}{s_{12} \, s_{1\four}}
 + \frac{d_{34\x 1\x 2}^{(0)}}{s_{12} \, s_{2\five}}
 + \frac{d_{34\x 2\x 1}^{(0)}}{s_{12} \, s_{1\five}} \nn \\ 
 &+& 2\, \frac{d_{1\x 2\x 3}^{(0)}}{s_{12} \, s_{23}} 
 + 2\,  \frac{d_{3\x 1\x 2}^{(0)}}{s_{12} \, s_{13}} \,.
\end{eqnarray}

\subsubsection{Boxes}

All the box coefficients are computed using unitarity cuts and subsequently simplified using
the analytic reconstruction techniques discussed in section~\ref{sec:advancements}. 
Although one might expect all four helicities to be required for each of the five boxes,
symmetry relations allow this number to be reduced.  We choose
to use the basic set shown in Table~\ref{allboxes}.  Other helicity combinations, or momentum
configurations, can be obtained through permutations of these.

\begin{table}
\begin{center}
\begin{tabular}{c|llll}
Configuration & \multicolumn{4}{c}{Helicities} \\
\hline
$1\x2\x3$  & $1^- ~ 2^- ~ 3^-$& $1^- ~ 2^- ~ 3^+$& $1^- ~ 2^+ ~ 3^-$& \\ 
$1\x2\x4$  & $1^- ~ 2^- ~ 3^-$& $1^- ~ 2^- ~ 3^+$& $1^- ~ 2^+ ~ 3^-$& $1^+ ~ 2^- ~ 3^-$ \\
$1\x4\x2$  & $1^- ~ 2^- ~ 3^-$& $1^- ~ 2^- ~ 3^+$& $1^- ~ 2^+ ~ 3^-$& \\
$1\x4\x23$ & $1^- ~ 2^- ~ 3^-$& $1^- ~ 2^- ~ 3^+$& $1^+ ~ 2^- ~ 3^-$& \\
$4\x1\x23$ & $1^- ~ 2^- ~ 3^-$& $1^- ~ 2^- ~ 3^+$& $1^+ ~ 2^- ~ 3^-$& 
\end{tabular}
\caption{All box coefficient functions needed in the calculation of the $0 \to gggHH$ amplitude.
\label{allboxes}}
\end{center}
\end{table}

\section{Analytic results for the process $0\to gggHH$}
\label{sec:analytic}
We now give detailed analytic results for the contributions to the process $0\to gggHH$ that do not involve the triple Higgs coupling.

\subsection{Scalar pentagons reduced to boxes}
\label{sec:pentbox}
In the process $0 \to gggHH$ we encounter for the first time pentagon integrals, so we now discuss the treatment of
such scalar pentagon integrals, which will be useful in the following.
In four dimensions the scalar pentagon integral can be reduced to a sum of the five box integrals obtained by removing
one propagator~\cite{Melrose:1965kb,vanNeerven:1983vr,Bern:1993kr}. This decomposition is detailed for the two pertinent
cases below.

\subsubsection{Case 1: Adjacent Higgs bosons}
For the pentagon scalar integral
with two adjacent Higgs bosons we have,
\begin{eqnarray} \label{pentagonreductionadjacent}
E_{0}(p_1,p_2,p_3,p_4;m)&=& 
 \mC^{1\x2\x3\x4}_1\,D_0(p_2,p_3,p_4;m)
+\mC^{1\x2\x3\x4}_2\,D_0(p_{12},p_3,p_4;m) \nn \\
&+&\mC^{1\x2\x3\x4}_3\,D_0(p_1,p_{23},p_4;m) \nn \\
&+&\mC^{1\x2\x3\x4}_4\,D_0(p_1,p_2,p_{34};m)
+\mC^{1\x2\x3\x4}_5\,D_0(p_1,p_2,p_3;m)\, .
\end{eqnarray}
In terms of the Cayley matrix the reduction coefficients are given by
\begin{equation} \label{Sinverse_equation}
\mC_i=-\frac{1}{2} \sum_{j} S^{-1}_{ij} \, .
\end{equation}The Cayley matrix,
$\left[S^{1\x2\x3\x4}\right]_{ij}=[m^2-\frac{1}{2}(q_{i-1}-q_{j-1})^2]$ 
where $q_i$ is the offset (affine) momentum is given by
\renewcommand{\baselinestretch}{1.3}
\begin{equation}
\label{Cayley1234}
  S^{1\x2\x3\x4}=
    \begin{pmatrix}
          m^2                   &  m^2                  &  m^2-\frac{1}{2}s_{12}&  m^2-\frac{1}{2}s_{\four\five} &  m^2-\frac{1}{2} \mh^2 \\
          m^2                   &  m^2                  &  m^2                  &  m^2-\frac{1}{2}s_{23} &  m^2-\frac{1}{2}s_{1\five} \\
          m^2-\frac{1}{2}s_{12} &  m^2                  &  m^2                  &  m^2                   &  m^2-\frac{1}{2}s_{3\four} \\
          m^2-\frac{1}{2}s_{\four\five} &  m^2-\frac{1}{2}s_{23}&  m^2                  &  m^2                   &  m^2-\frac{1}{2}\mh^2  \\
          m^2-\frac{1}{2} \mh^2 &  m^2-\frac{1}{2}s_{1\five}&  m^2-\frac{1}{2}s_{3\four}&  m^2-\frac{1}{2}\mh^2  &  m^2    
\end{pmatrix} \, .
\end{equation}
\renewcommand{\baselinestretch}{1}
Explicit forms for the pentagon reduction coefficients, $\mC^{1\x2\x3\x4}_i$ are (with $s_{ij}=(p_i+p_j)^2$),
\allowdisplaybreaks
\begin{eqnarray}
\label{cred1}
\mC^{1\x2\x3\x4}_1&=& -\frac{1}{32\; |S^{1\x2\x3\x4}|}\, \Big[s_{23}\, \big(s_{3\four}\, (s_{1\five}\, s_{\four\five}+s_{23}\, s_{3\four}-s_{3\four}\, s_{\four\five}-s_{12}\, s_{23}+s_{12}\, s_{1\five})\nn\\
  &+&\mh^2\, (s_{12}\, s_{3\four}-s_{23}\, s_{3\four}-2\, s_{12}\, s_{1\five})\big)\Big] \nn \\
&\equiv & -s_{23}\frac{2\spbab3.\four.\five.1\spaba1.\five.\four.3-s_{3\four}([13]\spaba1.\five.\four.3-\spa1.3\spbab3.\four.\five.1) }{32\; |S^{1\x2\x3\x4}|} \,,\\
\label{cred2}
\mC^{1\x2\x3\x4}_2&=& -\frac{1}{32\; |S^{1\x2\x3\x4}|}\, \Big[
     s_{3\four}s_{\four\five}\, \big(s_{3\four}\, s_{\four\five}-s_{1\five}\, s_{\four\five}-s_{23}\, s_{3\four}+s_{12}\, s_{23}+s_{12}\, s_{1\five}\big)\nn\\
     &+&\mh^2\, \big(s_{\four\five}\, (s_{23}\, s_{3\four}+s_{12}\, s_{1\five}-2\, s_{12}\, s_{3\four})-s_{12}\, (s_{23}\, s_{3\four}+s_{12}\, s_{23}+s_{12}\, s_{1\five})\big)\nn \\
     &+&\mh^4\, s_{12}\, (s_{23}+s_{12})\Big]  \\
&\equiv& -\frac{2\spab3.(1+2).3 \spaba1.\five.\four.3\spbab1.\five.\four.3-\spabab3.\four.\five.(1+2).3\,([13]\spaba1.\five.\four.3-\spa1.3\spbab3.\four.\five.1)}{32\; |S^{1\x2\x3\x4}|}\,, \nn \\
\label{cred3}
\mC^{1\x2\x3\x4}_3&=& -\frac{1}{32\; |S^{1\x2\x3\x4}|}\, \Big[s_{1\five}\, s_{\four\five}\, \big(s_{1\five}\, s_{\four\five}+s_{23}\, s_{3\four}
  +s_{12}\, s_{23}-s_{3\four}\, s_{\four\five}-s_{12}\, s_{1\five})\nn \\
  &+&\mh^2\, \big(s_{23}\, (s_{3\four}\, s_{\four\five}-s_{23}\, s_{3\four}-s_{12}\, s_{23})+s_{1\five}\, (s_{12}\, s_{\four\five}-s_{12}\, s_{23}-2\, s_{23}\, s_{\four\five})\big)\nn \\
  &+&\mh^4\, s_{23}\, (s_{23}+s_{12})\Big] \nn \\
 &\equiv& \mC^{1\x2\x3\x4}_2 (1 \leftrightarrow 3, \four \leftrightarrow \five)\,, \\
\label{cred4}
\mC^{1\x2\x3\x4}_4&=& -\frac{1}{32\; |S^{1\x2\x3\x4}|}\, \Big[s_{12}\, \big(s_{1\five}\, (s_{3\four}\, s_{\four\five}-s_{1\five}\, s_{\four\five}+s_{23}\, s_{3\four}-s_{12}\, s_{23}+s_{12}\, s_{1\five})\nn \\
       &+&\mh^2\, (s_{1\five}\, s_{23}-s_{12}\, s_{1\five}-2\, s_{23}\, s_{3\four})\big)\Big]\nn \\
&\equiv& \mC^{1\x2\x3\x4}_1 (1 \leftrightarrow 3, \four \leftrightarrow \five)\,, \\
\label{cred5}
\mC^{1\x2\x3\x4}_5&=& -\frac{1}{32\; |S^{1\x2\x3\x4}|}\, s_{12}\, s_{23}\, \Big[s_{3\four}\, s_{\four\five}+s_{1\five}\, s_{\four\five}-s_{23}\, s_{3\four}+s_{12}\, s_{23}-s_{12}\, s_{1\five}\nn\\
  &-&\mh^2\, (s_{12}+s_{23})\Big] \nn \\
  &\equiv& -s_{12}s_{23}\frac{[13] \spaba1.\five.\four.3-\spa1.3 \spbab3.\four.\five.1}{32\; |S^{1\x2\x3\x4}|} \,.
\end{eqnarray}
$\mC^{1\x2\x3\x4}_5$ is unchanged under $(1 \leftrightarrow 3, \four \leftrightarrow \five)$ exchange.
The alternative spinor expressions given in eq.~(\ref{cred1}-\ref{cred5}) have the merit that partial cancellations
which occur in the $m \to 0$ limit are made manifest, (see eq.~(\ref{Sdef2}) below).

The factor $|S^{1\x2\x3\x4}|$ is the determinant of the Cayley matrix,
eq.~(\ref{Cayley1234}). It can be written as,
\begin{equation}
\label{Sdef1}
16\, |S^{1\x2\x3\x4}|=-s_{12}\,s_{23}\,
 \Big(s_{1\five}\,s_{3\four}\,s_{\four\five}-\mh^2 \, (s_{23}\, s_{3\four}+s_{12}\, s_{1\five})\Big)
 +16 m^2\,\Delta(p_1,p_2,p_3,p_4)\,,
\end{equation}
where
\begin{eqnarray}
\Delta(p_1,p_2,p_3,p_4) &=& (p_1. p_2\; p_3. p_4-p_1. p_3\; p_2. p_4-p_1. p_4\; p_2. p_3)^2 
+2\; p_1. p_3\; p_2. p_3\; (p_1. p_2\; \mh^2-2\; p_1. p_4\; p_2. p_4)\nn \\
                       & =&\frac{1}{16}  \, (\trfive)^2 \,,
\end{eqnarray}
where $\trfive$ has been defined in eq.~(\ref{eq:tr5def}).
This can be written as another useful relation,
\begin{eqnarray}
\label{Sdef2}
16\, |S^{1\x2\x3\x4}|&=& -s_{12} \spab1.\slsh{p}_5 \slsh{p}_4\slsh{p}_3.2 \; \spab2.\slsh{p}_3\slsh{p}_4\slsh{p}_5.1
+ m^2 \, \big(\trfive\big)^2  \nn \\
&=& -s_{12}s_{23}\spaba1.\five.\four.3\spbab3.\four.\five.1+ m^2 \, \big(\trfive\big)^2 \, .
\end{eqnarray}

\subsubsection{Case 2: Non-adjacent Higgs bosons}
For the case of a scalar pentagon integral with non-adjacent Higgs bosons,
we denote the coefficients for the reduction of the scalar pentagon to scalar
boxes by $\mCbar^{1\x2\x4\x3}_i$,
\begin{eqnarray} \label{pentagonreductionadjacent2}
E_{0}(p_1,p_2,p_4,p_3;m)&=& 
 \mCbar^{1\x2\x4\x3}_1\,D_0(p_2,p_4,p_3;m)
+\mCbar^{1\x2\x4\x3}_2\,D_0(p_{12},p_4,p_3;m) \nn \\
&+&\mCbar^{1\x2\x4\x3}_3\,D_0(p_1,p_{24},p_3;m) \nn \\
&+&\mCbar^{1\x2\x4\x3}_4\,D_0(p_1,p_2,p_{34};m)
+\mCbar^{1\x2\x4\x3}_5\,D_0(p_1,p_2,p_4;m)\, .
\end{eqnarray}
For the case where the Higgs boson are not adjacent the Cayley matrix is given by
\renewcommand{\baselinestretch}{1.3}
\begin{equation}
\label{Cayley1243}
  S^{1\x2\x4\x3}=
    \begin{pmatrix}
          m^2                   &  m^2                  &  m^2-\frac{1}{2}s_{12}&  m^2-\frac{1}{2}s_{3\five} &  m^2-\frac{1}{2} \mh^2 \\
          m^2                   &  m^2                  &  m^2                  &  m^2-\frac{1}{2}s_{2\four} &  m^2-\frac{1}{2}s_{1\five} \\
          m^2-\frac{1}{2}s_{12} &  m^2                  &  m^2                  &  m^2-\frac{1}{2} \mh^2 &  m^2-\frac{1}{2}s_{3\four} \\
          m^2-\frac{1}{2}s_{3\five} &  m^2-\frac{1}{2}s_{2\four}&  m^2-\frac{1}{2} \mh^2&  m^2                   &  m^2                   \\
          m^2-\frac{1}{2} \mh^2 &  m^2-\frac{1}{2}s_{1\five}&  m^2-\frac{1}{2}s_{3\four}&  m^2                   &  m^2
\end{pmatrix}\,.
\end{equation}
\renewcommand{\baselinestretch}{1}
The reduction coefficients are given as before, using eq.~(\ref{Sinverse_equation}),
\allowdisplaybreaks
\begin{eqnarray}
\label{cbarred1}
\mCbar^{1\x2\x4\x3}_1&=& -\frac{1}{32\; |S^{1\x2\x4\x3}|}\,\Big[
(s_{2\four}\, s_{3\four}-\mh^2\, s_{1\five})\, \big(s_{1\five}\, s_{3\five}+s_{12}\, s_{1\five}+s_{2\four}\, s_{3\four}-s_{12}\, s_{2\four}-s_{3\four}\, s_{3\five}\nn\\ 
  &-&\mh^2\, (s_{1\five}+s_{2\four})+\mh^4\big)\Big]\nn \\
&\equiv& - \spab2.\four.3 \spab3.\four.2\frac{2 \spab1.\five.3 \spab3.\five.1+ \spab3.\four.1 \spab1.\five.3+\spab1.\four.3 \spab3.\five.1}{32\; |S^{1\x2\x4\x3}|} \,,\\
\label{cbarred2}
\mCbar^{1\x2\x4\x3}_2&=& -\frac{1}{32\; |S^{1\x2\x4\x3}|}\,\Big[
s_{3\four}\, s_{3\five}\, (s_{3\four}\, s_{3\five}-s_{1\five}\, s_{3\five}-s_{2\four}\, s_{3\four}+s_{12}\, s_{2\four}+s_{12}\, s_{1\five}) \nn\\
          &+&\mh^2\, (s_{2\four}\, s_{3\four}\, s_{3\five}+s_{1\five}\, s_{3\four}\, s_{3\five}-2\, s_{12}\, s_{1\five}\, s_{3\five}-2\, s_{12}\, s_{2\four}\, s_{3\four})\nn\\
          &+&\mh^4\, (s_{12}\, s_{1\five}+s_{1\five}\, s_{3\five}+s_{2\four}\, s_{3\four}+s_{12}\, s_{2\four}-2\, s_{3\four}\, s_{3\five})
-\mh^6\, (s_{1\five}+s_{2\four})+\mh^8 \Big]\nn \\
&\equiv & -\frac{-2 \spab3.\five.1 \spab2.\four.3\spab1.\five.3 \spab3.\four.2+(\spab3.\five.2 \spab3.\four.1 \spab1.\five.3 \spab2.\four.3)+(\langle\rangle \leftrightarrow [])}{32\; |S^{1\x2\x4\x3}|} \,,\\
\label{cbarred3}
\mCbar^{1\x2\x4\x3}_3&=& -\frac{1}{32\; |S^{1\x2\x4\x3}|}\,\Big[
(s_{1\five}\, s_{3\five}-\mh^2\, s_{2\four})\, (s_{1\five}\, s_{3\five}+s_{2\four}\, s_{3\four}+s_{12}\, s_{2\four}-s_{12}\, s_{1\five}-s_{3\four}\, s_{3\five}\nn\\
  &-&\mh^2\, (s_{1\five}+s_{2\four})+\mh^4)\Big]\nn \\
&\equiv&\mCbar^{1\x2\x4\x3}_1(1\leftrightarrow 2, \four \leftrightarrow \five) \,,\\
\label{cbarred4}
\mCbar^{1\x2\x4\x3}_4&=& -\frac{1}{32\; |S^{1\x2\x4\x3}|}\,\Big[
  s_{12}\, \big(s_{1\five}\, (s_{3\four}\, s_{3\five}-s_{1\five}\, s_{3\five}+s_{2\four}\, s_{3\four}+s_{12}\, s_{1\five}-s_{12}\, s_{2\four})\nn\\
&+&\mh^2\, (s_{1\five}\, s_{2\four}-s_{1\five}^2-2\, s_{2\four}\, s_{3\four})+\mh^4\, s_{1\five}\big) \Big]\nn \\
&\equiv& \frac{s_{12}(\spab1.\five.2 \spab3.\five.1 \spab2.\four.3+\spab2.\five.1 \spab1.\five.3 \spab3.\four.2)}{32\; |S^{1\x2\x4\x3}|} \,, \\
\label{cbarred5}
\mCbar^{1\x2\x4\x3}_5&=& -\frac{1}{32\; |S^{1\x2\x4\x3}|}\,\Big[
s_{12}\, \big(s_{2\four}\, (s_{3\four}\, s_{3\five}+s_{1\five}\, s_{3\five}-s_{12}\, s_{1\five}-s_{2\four}\, s_{3\four}+s_{12}\, s_{2\four})\nn\\
&+&\mh^2\, (s_{1\five}\, s_{2\four}-s_{2\four}^2-2\, s_{1\five}\, s_{3\five})+\mh^4\, s_{2\four}\big) \Big]\nn\\
&\equiv&\mCbar^{1\x2\x4\x3}_4(1\leftrightarrow 2, \four \leftrightarrow \five)\,.
\end{eqnarray}
$\mCbar^{1\x2\x4\x3}_2$ is symmetric under $(1 \leftrightarrow 2, \four \leftrightarrow \five)$.
For the non-adjacent Higgs boson case, the determinant of the Cayley matrix is,
\begin{equation}
16 |S^{1\x2\x4\x3}| =  -s_{12}\,(s_{2\four}\,s_{3\four}-\mh^2\,s_{1\five})
\,(s_{1\five}\,s_{3\five}-\mh^2\,s_{2\four}) + 16 m^2 \, \Delta(p_1,p_2,p_3,p_4)\,,
\end{equation}
or equivalently,
\begin{eqnarray}\label{SdefNonAdjacent}
  16 |S^{1\x2\x4\x3}| &=& -s_{12}\,\spab1.\slsh{p}_5\slsh{p}_3\slsh{p}_4.2 \; \spab2.\slsh{p}_4\slsh{p}_3\slsh{p}_5.1 + m^2 \, \big(\trfive\big)^2 \nn \\
  &=& -s_{12} \, \spab1.\five.3 \, \spab3.\five.1 \, \spab3.\four.2 \, \spab2.\four.3 \, + \, m^2 \, \big(\trfive\big)^2\,.
\end{eqnarray}

\subsection{$g^-g^-g^-HH$}

\subsubsection{Effective pentagons}

We will write the box coefficients ($d$) in terms of a combination of
effective pentagon ($\hat e$) and box ($\hat d$) coefficients. We begin
by specifying the effective pentagon coefficients. 

The effective pentagon coefficient (for adjacent Higgs bosons) is given by,
\begin{eqnarray}
\label{eq:Eff1x2x3x4mmm}
\hat{e}_{1\x2\x3\x4}(1^-,2^-,3^-) &=& \frac{m^2 s_{12} s_{23}}{4 \, \trfive} (8 m^2 - s_{\four\five} - 2 \mh^2) \spb1.3\*\spaba1.\five.\four.3 \,.
\end{eqnarray}

The effective pentagon coefficient
$\hat{e}_{1\x2\x4\x3}(1^-,2^-,3^-)$ (appropriate for the case where the Higgs bosons are not adjacent) is,
\begin{eqnarray}
  \hat{e}_{1\x2\x4\x3}(1^-,2^-,3^-) &=& \frac{m^2}{4} \spa1.2\spb1.3\spb2.3 \nn\\
  &\times&\Biggl\{ \frac{\spa1.2\spab3.\five.1\spab3.\four.2}{\trfive} (8m^2-s_{\four\five}-2\mh^2) 
  + \spaba3.\four.\five.3 \Biggr\}\,.
\end{eqnarray}
Note that $\hat e_{1\x2\x4\x3}(1^-,2^-,3^-)$ is manifestly symmetric under $(1\leftrightarrow 2, \four \leftrightarrow \five)$.
\subsubsection{Boxes}
The box coefficient is written in terms of an effective pentagon
coefficient ($\hat e$) plus a remainder term, $\hat{d}_{1\x2\x3}(1^-,2^-,3^-)$ as,
\begin{eqnarray}
d_{1\x2\x3}(1^-,2^-,3^-) &=& \Biggl\{ \frac{1}{\spb1.2 \spb2.3 \spb3.1}
 \Bigl[ \mC^{1\x2\x3\x4}_5 \, \hat e_{1\x2\x3\x4}(1^-,2^-,3^-) \Bigr] \Biggr\} 
 \nn \\
 && + \Bigg\{4 \leftrightarrow 5 \Bigg\} +\hat{d}_{1\x2\x3}(1^-,2^-,3^-) \,,
\end{eqnarray}
where the remainder term is
\begin{equation}
\hat{d}_{1\x2\x3}(1^-,2^-,3^-)=\frac{m^2}{2} \frac{\spa1.2 \spa2.3}{\spb3.1} \,.
\end{equation}
The reduction factor $\mC^{1\x2\x3\x4}_5$ is given in section~\ref{sec:pentbox}.

In a similar way we can write,
\begin{eqnarray}
d_{1\x2\x4}(1^-,2^-,3^-) &=& \frac{1}{\spb1.2 \spb2.3 \spb3.1}
\Bigl[ \mC^{3\x1\x2\x4}_1 \, \hat e_{1\x2\x3\x4}(3^-,1^-,2^-)
      +\mCbar^{1\x2\x4\x3}_5 \, \hat e_{1\x2\x4\x3}(1^-,2^-,3^-) \Bigr] \,,\nn \\
\end{eqnarray}
where in this case there is no remainder term,
\begin{equation}
\hat{d}_{1\x2\x4}(1^-,2^-,3^-)=0 \,.
\end{equation}
The effective pentagon coefficient $\hat e_{1\x2\x3\x4}(3^-,1^-,2^-)$ is obtained by simply permuting the
arguments in the defining eq.~(\ref{eq:Eff1x2x3x4mmm}), noting also that $\trfive$ 
does not flip sign under this even permutation.  The reduction coefficients are obtained similarly.

The remaining box coefficients are given by,
\begin{eqnarray}
  d_{1\x4\x2}(1^-,2^-,3^-) &=& \frac{1}{\spb1.2 \spb2.3 \spb3.1}
\Bigl[ \mCbar^{3\x2\x4\x1}_1 \, \hat e_{1\x2\x4\x3}(3^-,2^-,1^-)
      +\mCbar^{3\x1\x4\x2}_1 \, \hat e_{1\x2\x4\x3}(3^-,1^-,2^-)\Bigr] \nn \\
      &+&\hat{d}_{1\x4\x2}(1^-,2^-,3^-)\,,\\
\hat{d}_{1\x4\x2}(1^-,2^-,3^-)&=&\frac{(4m^2-\mh^2)(s_{1\four}s_{2\four}-\mh^2s_{3\five}) }{4 \spb1.2 \spb2.3 \spb3.1} \,,
\end{eqnarray}
and,
\begin{eqnarray}
  d_{4\x1\x23}(1^-,2^-,3^-) &=& \frac{1}{\spb1.2 \spb2.3 \spb3.1}
\Bigl[ \mC^{3\x2\x1\x4}_2 \, \hat e_{1\x2\x3\x4}(3^-,2^-,1^-)
      +\mC^{2\x3\x1\x4}_2 \, \hat e_{1\x2\x3\x4}(2^-,3^-,1^-)\Bigr] \nn \\
&+& \hat{d}_{4\x1\x23}(1^-,2^-,3^-) \\
\hat{d}_{4\x1\x23}(1^-,2^-,3^-)&=& \frac{\spa2.3\*\spaba1.\five.\four.1 }{\spb2.3} \frac{m^2}{2\*\trfive}\*(s_{\four\five}+2\*\mh^2-8\*m^2)\,,
\end{eqnarray}
and,
\begin{eqnarray}
  d_{1\x4\x23}(1^-,2^-,3^-) &=& \frac{1}{\spb1.2 \spb2.3 \spb3.1}
\Bigl[ \mCbar^{3\x2\x4\x1}_2 \, \hat e_{1\x2\x4\x3}(3^-,2^-,1^-)
      +\mCbar^{2\x3\x4\x1}_2 \, \hat e_{1\x2\x4\x3}(2^-,3^-,1^-) \nn \\
      &+& \hat{d}_{1\x4\x23}(1^-,2^-,3^-)\,.
\end{eqnarray}
Note that these last two boxes have the same remainder contribution.
\beq
\hat{d}_{1\x4\x23}(1^-,2^-,3^-)=\hat{d}_{4\x1\x23}(1^-,2^-,3^-)\,.
\eeq

This fully specifies the five integrals that enter the
basis set indicated in Table~\ref{allboxes}.  The remainder are related by,
\begin{eqnarray}
        d_{4\x1\x2}(1^-,2^-,3^-) &=& -d_{1\x2\x4}(2^-,1^-,3^-) \nn \\
        d_{34\x1\x2}(1^-,2^-,3^-) &=& d_{1\x2\x4}(1^-,2^-,3^-) \{4 \leftrightarrow 5\}  \nn \\
        d_{34\x2\x1}(1^-,2^-,3^-) &=& -d_{1\x2\x4}(2^-,1^-,3^-) \{4 \leftrightarrow 5\}  \nn \\
        d_{2\x34\x1}(1^-,2^-,3^-) &=& d_{1\x4\x2}(1^-,2^-,3^-) \{4 \leftrightarrow 5\}  \nn \\
        d_{1\x23\x4}(1^-,2^-,3^-) &=& d_{4\x1\x23}(1^-,2^-,3^-) \{4 \leftrightarrow 5\} \,,
\end{eqnarray}
with the full set obtained by performing cyclic permutations of $(1,2,3)$.

\subsubsection{Triangles}
The following triangle coefficients are all zero,
\begin{equation}
  c_{4\x12}(1^-,2^-,3^-) =0\,,\;\;\; c_{4\x123}(1^-,2^-,3^-) =0\,,\;\;\; c_{1\x2}(1^-,2^-,3^-) =0\, ,
\end{equation}
whereas the following two triangle coefficients only have contributions at order $m^2$
\begin{eqnarray}
c_{3\x12}^{(0)}(1^-,2^-,3^-) &=0\,,\;\;\;  c_{3\x12}^{(2)}(1^-,2^-,3^-) &= \frac{(s_{13}+s_{23})}{\spb1.2\,\spb2.3\,\spb3.1} \,,\\
 c_{1\x4}^{(0)}(1^-,2^-,3^-) &=0\,,\;\;\;  c_{1\x4}^{(2)}(1^-,2^-,3^-)&= -2 \frac{\spab1.\four.1}{\spb1.2\, \spb2.3\, \spb3.1} \,.
\end{eqnarray}
\subsubsection{Bubbles and rational terms}
As discussed in section~\ref{sec:bubbles}, for each helicity configuration
we need only give results for a single bubble coefficient.  In this case it vanishes,
\begin{equation}
  b_{12}(1^-,2^-,3^-) = 0\, .
\end{equation}
The rational term is
\begin{equation}
  R(1^-,2^-,3^-)=\frac{s_{12}+s_{23}+s_{31}}{\spb1.2 \, \spb2.3 \, \spb3.1} \, .
\end{equation}

 \subsection{$g^-g^-g^+HH$}

\subsubsection{Effective pentagons}
\label{sec:pentagonsmmp}

Turning now to the $1^- 2^- 3^+$ helicity, the first effective pentagon coefficient is,
\begin{eqnarray}
  \hat{e}_{1\x2\x3\x4}(1^-,2^-,3^+) = - \frac{m^2}{4} \spa1.3\spa2.3\spb1.2 &\Biggl\{& \frac{\spa1.2\spb2.3\spbab1.\five.\four.3}{\trfive} (s_{\four\five}-2s_{12}-2\mh^2+8m^2) \nn \\
  && + \spbab3.\five.\four.3 \Biggr\}\,.
\end{eqnarray}
The second effective pentagon coefficient is,
\begin{eqnarray}
  \hat{e}_{1\x2\x4\x3}(1^-,2^-,3^+) = - \frac{m^2}{4\*\trfive} \,  
  \spa1.3\spa2.3\spb1.2^2\*\spab1.\five.3\spab2.\four.3\*(s_{\four\five}-2s_{12}-2\mh^2+8\*m^2)\,.
\end{eqnarray}

The basis set specified in Table~\ref{allboxes} requires us to also define
the effective pentagon coefficients for the $(1^-,2^+,3^-)$ configuration.
The first of these, $\hat{e}_{1\x2\x3\x4}(1^-,2^+,3^-)$,
is symmetric under $(1\leftrightarrow 3, \four \leftrightarrow
\five)$.
This effective pentagon coefficient reads,
\begin{eqnarray}
  &&\hat{e}_{1\x2\x3\x4}(1^-,2^+,3^-) =
  \Bigg[\frac{s_{12}^2s_{23}\spaba1.\five.\four.3(\spabab3.\four.\five.(1+2).3+4m^2\spab3.(1+2).3)}{4\spa1.3\trfive}
    \nn \\ &&
    \;+\frac{1}{4}s_{12}m^2(s_{23}(s_{1\five}-s_{23})-\spa1.2[23]\spab3.\five.1)\Bigg]
  + \Bigg[\,\Bigg]_{1\leftrightarrow3,\,\four\leftrightarrow\five} \\ &&
  +m^2s_{12}s_{23}\Bigg([13]\spaba1.\five.\four.3\frac{3s_{123}-2s_{13}+2\mh^2-4m^2}{4\trfive}
  - m^2\frac{\spa1.3[1|\five|\four|3]}{\trfive} - 3m^2 \Bigg) \nn
\end{eqnarray}

The last effective pentagon is $\hat{e}_{1\x2\x4\x3}(1^-,2^+,3^-)$,
\begin{eqnarray}
  &&\hat{e}_{1\x2\x4\x3}(1^-,2^+,3^-) =
   \frac{m^2}{4} \spa1.2 \left(\begin{array}{c} \kern-33mm [23]\spab3.\five.1(s_{12}-s_{2\four}-\mh^2+8m^2)  \\
    \kern10mm -[12]\spab3.\five.3(\spab2.(3+\four).2-8\mh^2)+[13]\spab3.\four.2\spab2.\five.2\end{array}\right) \nn \\
  &+& \frac{m^2}{4} s_{12} \spab1.\five.3 \frac{\left(\begin{array}{c}\spa2.3[13]\spab3.\four.2(s_{123}-2s_{13}-2\mh^2+8m^2) \\
      \qquad-8\spab3.\five.3([12]\spa2.3\mh^2+\spab3.\five.1s_{12}) \end{array}\right)}{\trfive} \nn \\
  &+& \frac{1}{4}s_{12}\spa1.2\spab3.\five.1\spab3.\four.2\spab1.\five.3\Bigg[\frac{1}{\spa1.3}+\frac{[13](s_{2\four}+s_{3\five}-8m^2)+[12]\spab2.\five.3}{\trfive}\Bigg] \nn \\
  &+& \frac{1}{4}s_{12}\spa1.2\spab3.\five.1\spab3.\five.3\frac{\spab1.\five.3\spab3.\four.2(s_{123}-2\mh^2)-8m^2\spab1.\five.2\spab3.\five.3}{\spa1.3\trfive} \,.
\end{eqnarray}

\subsubsection{Boxes}

The first box coefficient can be written in terms of effective pentagons as,
\begin{eqnarray}
d_{1\x2\x3}(1^-,2^-,3^+) &=& \Biggl\{ \frac{\spa1.2}{\spb1.2^2 \spa2.3 \spa1.3}
 \mC^{1\x2\x3\x4}_5 \, \hat e_{1\x2\x3\x4}(1^-,2^-,3^+)\Biggr\} 
  + \Bigg\{4 \leftrightarrow 5 \Bigg\} \nn \\
 &+& \hat{d}_{1\x2\x3}(1^-,2^-,3^+), \label{eq:d123mmpgggHH}\\
 \hat{d}_{1\x2\x3}(1^-,2^-,3^+)&=&  \frac{m^2}{2} \frac{\spa1.2^2 \spb2.3 }{\spb1.2 \spa1.3}\,.
\end{eqnarray}
The next box coefficient again has no remainder,
\begin{eqnarray}
d_{1\x2\x4}(1^-,2^-,3^+) &=& \frac{\spa1.2}{\spb1.2^2 \spa2.3 \spa1.3} \nn \\
&\times & \Bigg[
  \mC^{3\x1\x2\x4}_1 \, \hat e_{1\x2\x3\x4}(3^+,1^-,2^-) 
 + \mCbar^{1\x2\x4\x3}_5 \, \hat e_{1\x2\x4\x3}(1^-,2^-,3^+)\Bigg]\,.
\end{eqnarray}
The next box coefficient can be decomposed as,
\begin{eqnarray}
  d_{1\x4\x2}(1^-,2^-,3^+)  \, & = & \, \frac{\spa1.2}{\spb1.2^2 \spa2.3 \spa1.3} \, \Bigl[ \mCbar^{3\x2\x4\x1}_1 \, \hat e_{1\x2\x4\x3}(3^+,2^-,1^-) + \nn \\
      && \kern26.5mm \mCbar^{3\x1\x4\x2}_1 \, \hat e_{1\x2\x4\x3}(3^+,1^-,2^-) \Bigr] \nn \\
      && \; + \; \hat d_{1\x4\x2}(1^-,2^-,3^+)  \, .
\end{eqnarray}
The remainder is anti-symmetric under the exchange $1\leftrightarrow 2$,
\begin{eqnarray}
&& \hat d_{1\x4\x2}(1^-,2^-,3^+) =
 \frac{\spa1.2\spab1.\four.2\spab2.\four.1}{4[12]} \nn \\
 &\times& \Bigg\{
 \frac{s_{13}+s_{23}-2\mh^2+8m^2}{\spa1.3\spa2.3} \Bigg[
   \frac{\spab1.\five.3\spa2.3+\spab2.\five.3\spa1.3}{\trfive} +
   \frac{1}{2[12]} \Bigg] +\frac{[3|\four|\five|3]}{\trfive} \Bigg\}
\end{eqnarray}
The next box coefficient is,
\begin{eqnarray}
  d_{4\x1\x23}(1^-,2^-,3^+) \, & = & \, \frac{\spa1.2}{\spb1.2^2 \spa2.3 \spa1.3} \nn\\
  &\times& \Bigl[ \mC^{3\x2\x1\x4}_2 \, \hat e_{1\x2\x3\x4}(3^+,2^-,1^-)  
      + \mC^{2\x3\x1\x4}_2 \, \hat e_{1\x2\x3\x4}(2^-,3^+,1^-) \Bigr] \nn \\
  &+& \hat d_{4\x1\x23}(1^-,2^-,3^+) \, ,
\end{eqnarray}
\begin{eqnarray}
&&  \hat d_{4\x1\x23}(1^-,2^-,3^+) =
  - \frac{[13]\spabab1.\four.\five.(2+3).1}{4[12][1|\four|\five|1]} \nn\\
  &\times& \Bigg[ \frac{[3|\five|\four|3]}{[23]} + [13](s_{123}-2\mh^2+8m^2) \Bigg( \frac{1}{2[12]} - \frac{[1|\four|\five|3]}{[23][1|\four|\five|1]}  \Bigg) \Bigg] \nn \\
  &+& \frac{\spa1.2}{4\trfive} \Big[ \mh^2\big(\spab1.\five.3\spab2.\four.3+\spab1.\four.3\spab2.\five.3\big)-\spab1.\four.3\spab2.\four.3\big(s_{123}-2\mh^2\big) \Big] \nn \\
  &+& \frac{s_{13}}{4[12]\trfive} \Big[ \spa1.2[13]\spab1.\five.3(s_{123}+\mh^2)-\spa1.2[23]\spab2.\five.3(\mh^2-8m^2)-2\spab2.\five.3\spab1.\five.3s_{123} \Big]  \nn \\
  &+& \frac{\spa1.2[13]}{4[12]\trfive} \Big[ s_{123}(\spab1.\five.2\spab2.\five.3+\spabab1.\four.{(2-3)}.\four.3)+\spa1.2[23](s_{13}s_{123}-2\spab3.\four.3\mh^2) \Big]
  \nn \\
  &+&[13][23]\frac{s_{123}-2\mh^2+8m^2}{4[12]^2} \Bigg[ s_{13} \frac{\spaba1.\four.\five.2+\spa1.2s_{3\five}}{\trfive} - \spa1.2 \Bigg] \nn \\
  && - m^2\frac{\spabab2.3.\four.\five.3-\spabab2.\five.\four.2.3}{2[12]\spa2.3} \Bigg[ (s_{123}-2\mh^2+8m^2) \Bigg( \frac{\spa1.2}{\trfive} + \frac{[13]}{[23][1|\four|\five|1]} \Bigg) - \frac{2\spa1.2s_{12}}{\trfive}\Bigg] \nn \\
    &+& \frac{s_{123}}{8[12]^2} \Bigg[ [23]\spab2.\four.3+[13]\spab1.\five.3 + \frac{\spab1.\five.2\spab2.\five.3-\spab1.\four.2 \spab2.\four.3}{\spa1.3} - \frac{[13]\spaba1.\four.\five.2}{\spa2.3} \Bigg] \nn \\
  &+& \spa1.2\frac{\spab1.\four.3(s_{123}-\mh^2)-\mh^2\spab1.\five.3}{8[12]\spa1.3} \,.
\end{eqnarray}
The last box coefficient is the most complicated,
\begin{eqnarray}
  d_{1\x4\x23}(1^-,2^-,3^+) \, & = & \, \frac{\spa1.2}{\spb1.2^2 \spa2.3 \spa1.3} \,
  \Bigl[ \big( \mCbar^{3\x2\x4\x1}_2 \, \hat e_{1\x2\x4\x3}(3^+,2^-,1^-) \big) + \big( \four \leftrightarrow \five \big) \Bigr] \nn \\
&& \; + \; \big( \hat d^{\,\text{unsym.}}_{1\x4\x23}(1^-,2^-,3^+) \big) + \big( \four \leftrightarrow \five \big) 
\end{eqnarray}

\begin{eqnarray}
&&\hat d^{\,\text{unsym.}}_{1\x4\x23}(1^-,2^-,3^+) =
  -\frac{[13]\spabab1.\five.(2+3).\four.1}{4[12]\spa2.3[1|\four|\five|1]} \nn \\
  &\times&\Bigg( \spab2.\four.1\frac{\spab2.\four.1\mh^2+4\spab2.\five.1m^2}{[1|\four|\five|1]} + \spab2.\five.1\frac{s_{123}-2\mh^2+8m^2}{2[12]} \Bigg) \nn \\
  &+& \frac{[13]m^2}{2[12][1|\four|\five|1]} \Bigg( \spa1.2\spab2.\five.1\frac{\mh^2-4m^2}{\spa2.3} - \spab1.\five.3\frac{[1|\four|\five|3]+4[13]m^2}{[23]} \Bigg)\nn \\
  &-&  \frac{\spa1.2\spab1.\four.1}{8[12]^2\spa1.3\spa2.3}(s_{123}-s_{12}-2\mh^2)(s_{123}-s_{23}-2\mh^2+8m^2+\spab1.\four.1) \nn \\
  &+& \frac{\spa1.2m^2}{2[12]^2\spa1.3\spa2.3}\big[s_{12}(2s_{3\five}-3s_{2\five})+4[12]\spa1.3\spab2.\four.3-2\spab1.\four.1(s_{2\five}+s_{3\five}-2\mh^2)\big] \nn \\
  &+& \frac{\spa1.2m^2}{2[12]\trfive} \Bigg[ \big(\spabab2.(2+3).\four.\five.3-\spabab2.\five.\four.(2+3).3\big) \frac{s_{12}-4m^2}{\spa2.3} - \spab1.\four.1\spab1.\five.3\frac{s_{123}-4s_{23}-2\mh^2}{\spa1.3} \Bigg] \nn \\
  &+& \frac{[23]\spab1.\four.1}{4[12]\trfive}\Bigg\{\spa1.2\spab2.\five.3(s_{123}-s_{12}-2\mh^2+s_{3\five}+2s_{2\five})+\spa1.3\spab2.\five.3^2 \nn \\
  &+&\spa1.2\spab2.\four.3\mh^2 - \spa1.2^2\frac{\spab3.\four.2 \spab2.\five.3+s_{123}(s_{123}-\mh^2)+2\spab3.\five.3\mh^2-\spab2.\five.2^2}{\spa1.3} \nn \\
  &+& \spab1.\four.1\spaba1.\five.\four.2\frac{s_{123}-2\mh^2+8m^2}{[12]\spa1.3} \Bigg\}
\end{eqnarray}

Since the above relationships involve permuting the arguments $1^-$, $2^-$
and $3^+$ they result in contributions from effective pentagon coefficients
with other helicity orderings.  These can be simply related to our
basis set by reading off the momenta in the opposite direction around the loop:
\begin{eqnarray}
\hat e_{1\x2\x3\x4}(3^+,1^-,2^-) &=& \hat e_{1\x2\x3\x4}(2^-,1^-,3^+) (4 \leftrightarrow 5) \nn \\
\hat e_{1\x2\x4\x3}(3^+,2^-,1^-) &=& \hat e_{1\x2\x4\x3}(2^-,3^+,1^-) (4 \leftrightarrow 5) \nn \\
\hat e_{1\x2\x4\x3}(3^+,1^-,2^-) &=& \hat e_{1\x2\x4\x3}(1^-,3^+,2^-) (4 \leftrightarrow 5) \nn \\
\hat e_{1\x2\x3\x4}(3^+,2^-,1^-) &=& \hat e_{1\x2\x3\x4}(1^-,2^-,3^+) (4 \leftrightarrow 5) \,.
\end{eqnarray}

This fully specifies the five integrals that enter the
basis set indicated in Table~\ref{allboxes}.  The remainder are related by,
\begin{eqnarray}
        d_{4\x1\x2}(1^-,2^-,3^+) &=& -d_{1\x2\x4}(2^-,1^-,3^+) \nn \\
        d_{34\x1\x2}(1^-,2^-,3^+) &=& d_{1\x2\x4}(1^-,2^-,3^+) \{4 \leftrightarrow 5\} \nn \\
        d_{34\x2\x1}(1^-,2^-,3^+) &=& -d_{1\x2\x4}(2^-,1^-,3^+) \{4 \leftrightarrow 5\} \nn \\
        d_{2\x34\x1}(1^-,2^-,3^+) &=& d_{1\x4\x2}(1^-,2^-,3^+) \{4 \leftrightarrow 5\} \nn \\
        d_{1\x23\x4}(1^-,2^-,3^+) &=& d_{4\x1\x23}(1^-,2^-,3^+) \{4 \leftrightarrow 5\} \,,
\end{eqnarray}
with the full set obtained by performing cyclic permutations of $(1,2,3)$.

\subsubsection{Triangles}
The following triangle coefficients are zero,
\begin{equation}
  c_{4\x12}(1^-,2^-,3^+) =0\,,\;\;\; c_{1\x2}(1^-,2^-,3^+) =0\, ,
\end{equation}
whereas the coefficients $c_{3\x12}(1^-,2^-,3^+)$ and $c_{1\x4}^{(2)}(1^-,2^-,3^+)$ only have a contribution at order $m^2$
\begin{equation}
  c_{3\x12}^{(0)}(1^-,2^-,3^+) =0\,,\;\;\;
  c_{3\x12}^{(2)}(1^-,2^-,3^+) = \frac{\spa1.2}{\spb1.2^2} \, \frac{(s_{13}+s_{23})}{\spa1.3\,\spa2.3}
\end{equation}
In addition for $c_{1\x4}(1^-,2^-,3^+)$ we have,
\begin{eqnarray}
  c_{1\x4}^{(0)}(1^-,2^-,3^+)&=&0\,,\nn\\
  c_{1\x4}^{(2)}(1^-,2^-,3^+)&=& -2 \frac{\spab1.\four.1}{\spb1.2\, \spa2.3\, \spb2.3}
   \big[ \frac{\spb1.3^2 \, \spab2.\four.1 \big(\spab1.\four.1-\spab2.3.2\big)}{\spbb1.\four.\five.1^2}\nn \\
  &-& \frac{\big(\spab3.\four.1 \, \spab2.\five.3\, \spb1.3+\spab2.\four.1 \, \spab2.\five.1\, \spb2.3\big)}{\spbb1.\four.\five.1\, \spab3.\four.1}\big]\,.
\end{eqnarray}
$c_{4\x123}(1^-,2^-,3^+)$ has contributions at both order $m^0$ and $m^2$,
\begin{eqnarray}
c_{4\x123}^{(0)}(1^-,2^-,3^+)&=&\frac{(s_{\four\five}-2\*\mh^2)}{8} c_{4\x123}^{(2)}(1^-,2^-,3^+)  \,, \label{eq:c4x123mmpxm0} \\
c_{4\x123}^{(2)}(1^-,2^-,3^+)&=&
  \Big\{2\*\frac{\spb1.3^3\*s_{\four\five}(s_{\four\five}-4\*\mh^2)}{\spb1.2\*\spb2.3\*\spbab1.\four.\five.1^2}\nn \\
- 4 \*\frac{\spb1.3\*(2\*\spb2.3\*\spab2.\four.3+\spab1.\four.3\*\spb1.3)}{\spb1.2\*\spb2.3\*\spbab1.\four.\five.1}
&-&2\*\frac{(s_{3\four}+\mh^2-s_{1\four}-s_{2\four})\*\spb1.3^2}{\spb1.2^2\*\spbab1.\four.\five.1} \Big\}
-\Big\{ 1\leftrightarrow 2 \Big\} \,. \label{eq:c4x123mmpxm2}
\end{eqnarray}
It is interesting to note that
$s_{\four\five}(s_{\four\five}-4\*\mh^2)$ is the factorized form of
the K\"{a}ll{\'e}n function $\Delta_{123|\four|\five}$ under the
constraint $m_\four=m_\five$. Without the constraint it is
unfactorizable. Furthermore, the second power of the pole $\spb1.2$ is
spurious. In appendix \ref{sec:spinor_decompositions} we show how to
eliminate it and the consequences this has on the other poles in light
of primary decompositions in the covariant spinor ring.

\subsubsection{Bubbles and rational terms}
All the bubble coefficients vanish,
\begin{equation}
  b_{12}(1^-,2^-,3^+) = 0 \,.
\end{equation}
The rational term is
\begin{equation}
  R(1^-,2^-,3^+)=-\frac{\spa1.2^2}{\spb1.2}\, \Big[
   \frac{1}{\spa1.3 \, \spa2.3}
  +\frac{\spb2.3}{\spa1.3 \, (s_{12}+s_{23})}
  +\frac{\spb1.3}{\spa2.3 \, (s_{12}+s_{13})}\Big]\,.
\end{equation}

\subsection{$g^-g^+g^-HH$}

\subsubsection{Effective pentagons}
Turning now to the $1^-2^+3^-$ helicity combination all the effective pentagon coefficients necessary for this amplitude
have already been introduced in section~\ref{sec:pentagonsmmp}.
\subsubsection{Boxes}
In terms of effective pentagons the first box is,
\begin{eqnarray}
  d_{1\x2\x3}(1^-,2^+,3^-) \, & = & \, \frac{\spa1.3}{\spb1.3^2\*\spa2.3\*\spa1.2} \,
  \Bigl[ \big(\mC^{1\x2\x3\x4}_5 \, \hat e_{1\x2\x3\x4}(1^-,2^+,3^-)\big) + \big(\four \leftrightarrow \five\big) \Bigr] \nn \\[2mm]
      && \; + \; \hat d_{1\x2\x3}(1^-,2^+,3^-) \, ,
\end{eqnarray}
\begin{eqnarray}
  \hat d_{1\x2\x3}(1^-,2^+,3^-) &=&
  \Bigg[\frac{\spa1.3[23] \spab3.\five.2s_{12}s_{123}}{4[13]\trfive}\Bigg]
  + \Bigg[\four \leftrightarrow \five \Bigg]
  - \Bigg[1 \leftrightarrow 3 \Bigg]
  - \Bigg[1 \leftrightarrow 3, \four \leftrightarrow \five \Bigg] \nn \\
  && + \frac{[12][23]s_{12}s_{23}}{2[13]^3}
  +\frac{[12]\spa1.3[23](s_{123}-6m^2)}{4[13]^2}
  -\frac{[12]\spa1.3^2[23]}{4[13]}
\end{eqnarray}

The second box is,
\begin{eqnarray}
  d_{1\x2\x4}(1^-,2^+,3^-)  \, & = & \, \frac{ \spa1.3}{\spb1.3^2\* \spa2.3\* \spa1.2} \,
  \Bigl[ \mC^{3\x1\x2\x4}_1 \, \hat e_{1\x2\x3\x4}(3^-,1^-,2^+) + \nn \\
      && \kern26mm \mCbar^{1\x2\x4\x3}_5 \, \hat e_{1\x2\x4\x3}(1^-,2^+,3^-) \Bigr] \nn \\
      && \; + \; \hat d_{1\x2\x4}(1^-,2^+,3^-) \, ,
\end{eqnarray}
\begin{eqnarray}
  \hat d_{1\x2\x4}(1^-,2^+,3^-) &=&
  \frac{s_{12}}{8\spa2.3[13]^2} \Bigg[  \spab3.\four.2(\mh^2-2s_{3\five})- \spab3.\five.2 s_{2\four} -   \spab3.\five.1\frac{s_{2\four}(s_{2\four}-8m^2)+\mh^4}{  \spab2.\four.1} \Bigg] \nn
  \\
  && + \frac{  \spab3.\five.1  \spab1.\four.2s_{12}\big[  \spab3.\four.2 \spab2.\five.3+ \spab3.\four.3 \mh^2-\spab3.(1+\five).3(s_{2\four}-8m^2)\big]}{4\spa2.3[13]^2\trfive} \quad\qquad
\end{eqnarray}

The last box we need to define is,

\begin{eqnarray}
  d_{1\x4\x2}(1^-,2^+,3^-)  \, & = & \, \frac{ \spa1.3}{ \spb1.3^2\* \spa2.3\* \spa1.2} \,
  \Bigl[ \mCbar^{3\x2\x4\x1}_1 \, \hat e_{1\x2\x4\x3}(3^-,2^+,1^-)  + \nn \\
      && \kern26mm \mCbar^{3\x1\x4\x2}_1 \, \hat e_{1\x2\x4\x3}(3^-,1^-,2^+) \Bigr] \nn \\
      && \; + \; \hat d_{1\x4\x2}(1^-,2^+,3^-) \, ,
\end{eqnarray}

\begin{eqnarray}
  \hat d_{1\x4\x2}(1^-,2^+,3^-) &=& \frac{ \spab1.\four.2}{8[13]^2} \Bigg( \frac{1}{\spa1.2} - 2 \frac{[23] \spab3.\five.1}{\trfive} \Bigg) \nn \\[2mm]
  && \times \Big[ \spab2.\four.1 \spab1.\five.2+ \spab1.\four.1 \mh^2-\big( \spab2.\four.2- \spab3.\five.3\big)\big(s_{2\four}-8m^2\big)\Big]
\end{eqnarray}

Again we can relate some of these pentagon coefficients to already-specified ones, after permutation:
\begin{eqnarray}
\hat e_{1\x2\x3\x4}(2^+,3^-,1^-) &=& e_{1\x2\x3\x4}(1^-,3^-,2^+) (4 \leftrightarrow 5) \\ 
\hat e_{1\x2\x4\x3}(2^+,3^-,1^-) &=& e_{1\x2\x4\x3}(3^-,2^+,1^-) (4 \leftrightarrow 5)
\end{eqnarray}

This fully specifies the three integrals that enter the
basis set indicated in Table~\ref{allboxes}. The remainder are related by,
\begin{eqnarray}
        d_{4\x1\x2}(1^-,2^+,3^-) &=& -d_{1\x2\x4}(2^+,1^-,3^-) \nn \\
        d_{34\x1\x2}(1^-,2^+,3^-) &=& d_{1\x2\x4}(1^-,2^+,3^-) \{4 \leftrightarrow 5\} \nn \\
        d_{34\x2\x1}(1^-,2^+,3^-) &=& -d_{1\x2\x4}(2^+,1^-,3^-) \{4 \leftrightarrow 5\} \nn \\
        d_{2\x34\x1}(1^-,2^+,3^-) &=& d_{1\x4\x2}(1^-,2^+,3^-) \{4 \leftrightarrow 5\} \nn \\
        d_{1\x23\x4}(1^-,2^+,3^-) &=& -d_{4\x1\x23}(1^-,3^-,2^+) \{4 \leftrightarrow 5\} \nn \\
        d_{4\x1\x23}(1^-,2^+,3^-) &=& -d_{4\x1\x23}(1^-,3^-,2^+) \nn \\
        d_{1\x4\x23}(1^-,2^+,3^-) &=& -d_{1\x4\x23}(1^-,3^-,2^+) \,,
\end{eqnarray}
with the full set obtained by performing cyclic permutations of $(1,2,3)$.
Note that two of these relations involve the box coefficients for the
$(+, -, -)$ configuration that will be given in the following section.

\subsubsection{Triangles}
The coefficient $c_{1\x2}(1^-,2^+,3^-)$ is very simple,
\begin{equation}
  c_{1\x2}(1^-,2^+,3^-)=-\frac{s_{12} \, \spb1.2\, \spb2.3}{2 \spb1.3^3}
\end{equation}
The following two triangle coefficients only have contributions at order $m^2$,
\begin{eqnarray}
  c_{3\x12}^{(0)}(1^-,2^+,3^-)&=&0 \\
  c_{3\x12}^{(2)}(1^-,2^+,3^-) &=& \Big[ 
  \frac{2 \, (s_{\four\five}-2 \mh^2) \, \spb2.3^3\, (s_{13}+s_{23})}{\spb1.2\, \spb1.3\, \spbb3.\four.\five.3^2}
      - \frac{2\, \spb2.3\,\spa1.3^2}{\spb1.3\,\spa1.2\, (s_{13}+s_{23})} \nn\\
     &+&  \frac{\spb2.3\,\spa1.3}{\spb1.3^2\,\spa1.2}
      -  \frac{\spa1.3^2}{\spb1.3\,\spa1.2\,\spa2.3}\Big]
\end{eqnarray}
and,
\begin{eqnarray}
  c_{1\x4}^{(0)}(1^-,2^+,3^-) &=& -c_{1\x4}^{(0)}(1^-,3^-,2^+)=0 \\
  c_{1\x4}^{(2)}(1^-,2^+,3^-) &=& -c_{1\x4}^{(2)}(1^-,3^-,2^+) \\
  &=& 2 \frac{\spab1.\four.1}{\spb1.3\, \spa3.2\, \spb3.2}
   \big[ \frac{\spb1.2^2 \, \spab3.\four.1 \big(\spab1.\four.1-\spab3.2.3\big)}{\spaa1.\five.\four.1^2}\nn \\
  &+& \frac{\big(\spab2.\four.1 \, \spab3.\five.2\, \spb1.2+\spab3.\four.1 \, \spab3.\five.1\, \spb3.2\big)}{\spaa1.\five.\four.1\, \spba1.\four.2}\big]  
\end{eqnarray}
The last triangle coefficient is simply related to one previously defined,
\begin{eqnarray}
  c_{4\x12}(1^-,2^+,3^-)&=&-c_{4\x12}(2^+,1^-,3^-) \,,\\
  c_{4\x123}(1^-,2^+,3^-)&=&c_{4\x123}(3^-,1^-,2^+) \,.
\end{eqnarray}
\subsubsection{Bubbles and rational terms}
The bubble coefficient $b_{12}(1^-,2^+,3^-)$ is given by,
\begin{equation}
  b_{12}(1^-,2^+,3^-)=-\frac{\spa1.3 \, \spb1.2}{\spa2.3\,\spb1.3^2}
  -\frac{\spa1.3^2 \, \spb1.2}{\spa2.3\,\spb1.3 (s_{13}+s_{23})}
  -\frac{\spa1.3^2 \, \spb1.2 \, \spb2.3}{\spb1.3 (s_{13}+s_{23})^2}
\end{equation}
The rational term is
\begin{equation}
  R(1^-,2^+,3^-)=-\frac{\spa1.3^2}{\spa1.2 \,\spa2.3\,\spb1.3}
  \Big[1-\frac{s_{12}}{(s_{12}+s_{13})}-\frac{s_{23}}{(s_{13}+s_{23})}\Big]
\end{equation}

 \subsection{$g^+g^-g^-HH$}

\subsubsection{Effective pentagons}

Turning now to the $1^+2^-3^-$ helicity combination, all the effective pentagon coefficients necessary for this amplitude
have already been introduced in section~\ref{sec:pentagonsmmp}.
\subsubsection{Boxes}
There is one independent box with two lightlike external lines,
\begin{eqnarray}
  d_{1\x2\x4}(1^+,2^-,3^-) \, & = & \, \frac{\spa2.3}{\spb2.3^2\*\spa1.3\*\spa1.2} \nn \\
   &\times& \Bigl[ \mC^{3\x1\x2\x4}_1 \, \hat e_{1\x2\x3\x4}(3^-,1^+,2^-)  + \mCbar^{1\x2\x4\x3}_5 \, \hat e_{1\x2\x4\x3}(1^+,2^-,3^+) \Bigr] \nn \\
      &+& \hat d_{1\x2\x4}(1^+,2^-,3^-)  \, ,
\end{eqnarray}
where the remainder $\hat d_{1\x2\x4}(1^+,2^-,3^-)$ reads,
\begin{eqnarray}
  &&\hat d_{1\x2\x4}(1^+,2^-,3^-) = -\frac{[12]s_{2\four}}{8\spa1.3 [23]^2} \Bigg\{
    \frac{\spa1.3 \big[s_{12}(s_{1\four}+\mh^2)+2\mh^2(s_{2\four}-\mh^2)\big]+8\spa1.2 \spab3.\five.2m^2}{\spab1.\four.2} \nn \\
    && + \spa2.3 (s_{13}+2s_{12}-2\mh^2) \Bigg\} +\frac{[12]\spab2.\four.1}{4[23]^2\trfive} \Bigg\{ - \spa1.2 [13]\Big[\spab3.\four.2(s_{1\five}-s_{3\five}+2\mh^2) \nn \\
    &&  +8\spab3.\five.2 m^2\Big] + \Big[ \big(s_{12}(\spab3.\five.3(s_{123}-2\mh^2)+(s_{3\five}-s_{3\four})\mh^2)\big)+\big(2 \leftrightarrow 3, \four \leftrightarrow \five\big) \Big] \nn \\
   && -\spa1.2  \spab3.\four.2 \frac{\spab3.\five.3 (s_{12}-s_{13}-2\mh^2+8m^2)+[12]\spa1.3 \spab2.\five.3}{\spa1.3 } \Bigg\}
\end{eqnarray}

The first box with only one lightlike external line is,
\begin{eqnarray}
  d_{4\x1\x23}(1^+,2^-,3^-) \, & = & \, \frac{\spa2.3}{\spb2.3^2\*\spa1.3\*\spa1.2} \nn \\
  &\times& \Bigl[ \mC^{3\x2\x1\x4}_2 \, \hat e_{1\x2\x3\x4}(3^-,2^-,1^+)+\mC^{2\x3\x1\x4}_2 \, \hat e_{1\x2\x3\x4}(2^-,3^-,1^+) \Bigr] \nn \\
      &+& \hat d_{4\x1\x23}(1^+,2^-,3^-) \, ,
\end{eqnarray}
where the effective box coefficient $\hat d_{4\x1\x23}(1^+,2^-,3^-)$ is given by
\begin{eqnarray}
  \hat d_{4\x1\x23}(1^+,2^-,3^-) = \frac{1}{2}m^2(s_{123}-2s_{23}-2\mh^2+8m^2)\frac{\spa2.3 }{[23]}\frac{[1|\four|\five|1]}{\trfive}\, .
\end{eqnarray}
It is manifestly $(\four \leftrightarrow \five)$ symmetric.

The other box with one lightlike external line has the same effective box contribution,
\begin{eqnarray}
  d_{1\x4\x23}(1^+,2^-,3^-) \, & = & \, \frac{\spa2.3}{\spb2.3^2\*\spa1.3\*\spa1.2} \nn \\
  &\times& \Bigl[ \mCbar^{3\x2\x4\x1}_2 \, \hat e_{1\x2\x4\x3}(3^-,2^-,1^+) +
        +\mCbar^{2\x3\x4\x1}_2 \, \hat e_{1\x2\x4\x3}(2^-,3^-,1^+) \Bigr] \nn \\
      &+& \hat d_{1\x4\x23}(1^+,2^-,3^-) \, ,
\end{eqnarray}
with
\begin{eqnarray}
  \hat d_{1\x4\x23}(1^+,2^-,3^-) = \hat d_{4\x1\x23}(1^+,2^-,3^-) \, .
\end{eqnarray}

This fully specifies the three integrals that enter the
basis set indicated in Table~\ref{allboxes}.  The remainder are related by,
\begin{eqnarray}
        d_{1\x2\x3}(1^+,2^-,3^-) &=& -d_{1\x2\x3}(3^-,2^-,1^+) \nn \\
        d_{4\x1\x2}(1^+,2^-,3^-) &=&-d_{1\x2\x4}(2^-,1^+,3^-) \nn \\
        d_{34\x1\x2}(1^+,2^-,3^-) &=& d_{1\x2\x4}(1^+,2^-,3^-) \{4 \leftrightarrow 5\}\nn \\
        d_{34\x2\x1}(1^+,2^-,3^-) &=& -d_{1\x2\x4}(2^-,1^+,3^-) \{4 \leftrightarrow 5\} \nn\\
        d_{1\x4\x2}(1^+,2^-,3^-) &=&  -d_{1\x4\x2}(2^-,1^+,3^-) \nn\\
        d_{2\x34\x1}(1^+,2^-,3^-) &=& -d_{1\x4\x2}(2^-,1^+,3^-) \{4 \leftrightarrow 5\} \nn \\
        d_{1\x23\x4}(1^+,2^-,3^-) &=& d_{4\x1\x23}(1^+,2^-,3^-) \{4 \leftrightarrow 5\} \,,
\end{eqnarray}
with the full set obtained by performing cyclic permutations of $(1,2,3)$.

\subsubsection{Triangles}
The simplest triangle coefficient is,
\begin{equation}
  c_{1\x2}(1^+,2^-,3^-)=-\frac{s_{12} \, \spb1.2 \, \spb1.3}{2 \spb2.3^3}\,.
\end{equation}
The order $m^2$ coefficients $c_{3\x12}^{(2)}(1^+,2^-,3^-)$ and
$c_{1\x4}^{(2)}(1^+,2^-,3^-)$ are given by, 
\begin{eqnarray}
  c_{3\x12}^{(2)}(1^+,2^-,3^-) &=& -c_{3\x12}^{(2)}(2^-,1^+,3^-) \\
 &=& -\Big[ 
  \frac{-2 \, (s_{\four\five}-2 \mh^2) \, \spb1.3^3\, (s_{13}+s_{23})}{\spb2.1\, \spb2.3\, \spbb3.\four.\five.3^2}
      - \frac{2\, \spb1.3\,\spa2.3^2}{\spb2.3\,\spa2.1\, (s_{13}+s_{23})} \nn\\
     &+&  \frac{\spb1.3\,\spa2.3}{\spb2.3^2\,\spa2.1}
      -  \frac{\spa2.3^2}{\spb2.3\,\spa2.1\,\spa1.3}\Big]\,,
\end{eqnarray}
and,
\begin{eqnarray}
  c_{1\x4}^{(2)}(1^+,2^-,3^-) &=& 2 \frac{\spab1.\four.1}{\spb2.3^2}
 \Big[
    \frac{ \spab3.\five.2}{\spa1.3\,\spab1.\four.2}-\frac{ \spab2.\five.3}{\spa1.2\,\spab1.\four.3}
    +\frac{\spa3.2}{\spa1.2\,\spa1.3}\Big]\,,
\end{eqnarray}
with the order $m^0$ pieces determined by the infrared relations
given in eqs.~(\ref{IR1}) and~(\ref{IR2}).

The remaining triangle is specified by,
\begin{eqnarray}\label{eq:c0x4x12pmmxm0}
c^{(0)}_{4\x12}(1^+,2^-,3^-)
&=&\frac{\spab2.\four.1
      \*(s_{2\four}-s_{1\four})}{4\*s_{12}\*\spbab3.\four.\five.3}
      \*\Bigg[\spb1.3\*\spa2.3 +
      \spa1.3\*\frac{\spb1.2\*\spab2.\five.3+\spb1.3\*(s_{3\five}-s_{2\four})}
      {\spab1.\four.2}\Bigg] \nn \\
&-&\frac{\spa2.3\*\Delta_{12|\four|3\five}
      \*(\spb1.2\*\spab2.\four.3-\spb1.3\*(s_{1\four}-\mh^2))}
      {4\*s_{12}\*\spab1.\four.2\*\spbab3.\four.\five.3}\nn \\
&+&\frac{\spab2.\four.1\*(s_{2\four}-s_{1\four})
      \*(s_{3\five}-s_{2\four})\*(s_{123}-2\*\mh^2)}
      {4\*s_{12}\*\spab1.\four.2\*\spbab3.\four.\five.3}
      -\frac{\spab2.\four.1^2\*(s_{12}-2\mh^2)}
      {2\*s_{12}\*\spbab3.\four.\five.3} \nn \\
&+&\frac{\spab2.\five.3\*\Delta_{12|\four|3\five}\*(s_{123}-2\*\mh^2)
      \*(\spb1.2\*\spab2.\five.3+\spb1.3
      \*(s_{3\five}-s_{2\four}))}
      {4\*s_{12}\*\spab1.\four.2\*\spbab3.\four.\five.3^2}\,,
\end{eqnarray}
\begin{eqnarray}\label{eq:c0x4x12pmmxm2}
  c^{(2)}_{4\x12}(1^+,2^-,3^-)&=&
  2 \spab2.\four.1 \*\frac{(s_{2\four}-s_{1\four})\*(s_{3\five}-s_{2\four})-2\*\spab1.\four.2 \spab2.\four.1}
    {\spab1.\four.2\* \spbab3.\four.\five.3 \* s_{12}} \nn \\
  &+&2\*\frac{\spab2.\five.3\*\Delta_{12|\four|3\five}\*(\spb1.2\*\spab2.\five.3+\spb1.3\*(s_{3\five}-s_{2\four}))}
    {\spab1.\four.2\* \spbab3.\four.\five.3^2 \* s_{12}}\,,
\end{eqnarray}
where the K\"{a}ll{\'e}n function $\Delta_{12|\four|3\five}$ was defined in eq.~\eqref{eq:delta12-4-35}
and
\begin{equation}
  c_{4\x123}(1^+,2^-,3^-)=c_{4\x123}(2^-,3^-,1^+) \,.
\end{equation}
It is evident that eq.~\eqref{eq:c0x4x12pmmxm2} could be reabsorbed
into eq.~\eqref{eq:c0x4x12pmmxm0} by adding $8m^2$ to the parentheses
involving $\mh^2$ in the last three fractions.

\subsubsection{Bubbles and rational terms}
The bubble coefficient is given by,
\begin{equation}
  b_{12}(1^+,2^-,3^-) = -\frac{\spa2.3 \spb1.2 \spb1.3}{\spb2.3}
        \Big[\frac{\spa2.3}{(s_{13}+s_{23})^2}
        -\frac{1}{\spb2.3 (s_{13}+s_{23})}\Big]
\end{equation}
The rational term is
\begin{equation}
  R(1^+,2^-,3^-)=-\frac{\spa2.3^2}{\spb2.3}\Big[
  \frac{1}{\spa1.2 \, \spa1.3}
  +\frac{\spb1.3 }{\spa1.2 \, (s_{13}+s_{23})}
  +\frac{\spb1.2 }{\spa1.3 \, (s_{12}+s_{23})}\Big]
\end{equation}

\section{Implementation of NLO $pp \to HH$ calculation}
The one-loop matrix elements that we have computed here have been cross-checked against
OpenLoops~\cite{Buccioni:2019sur}, finding full agreement.  Our analytic calculation
of the $0 \to gggHH$ process is approximately 90 times faster to evaluate than {\tt OpenLoops}, while
the simpler $0 \to \bar qqgHH$ amplitude is only 35 times quicker.
The implementation of the corresponding dipole subtraction terms in {\tt MCFM}~\cite{MCFM,Campbell:2015qma,Campbell:2019dru},
to complete the real radiation computation, is straightforward.

The 2-loop virtual matrix element contribution is implemented using {\tt hhgrid}~\cite{Heinrich:2017kxx},
a package that uses a grid to interpolate the two-loop result
(available from {\tt https://github.com/mppmu/hhgrid}).
The package provides a Fortran interface to the interpolating Python code, which we have linked
to {\tt MCFM}.\footnote{
We thank Stephen Jones for help producing grid files that can be loaded efficiently, so that it is
straightforward to run our calculations on multiple cores simultaneously.}

The {\tt hhgrid} code provides the value of ${\cal V}_{\rm fin}$, which is defined in terms of the
virtual contribution ${\cal V}_b$ --  the interference of the 2-loop and 1-loop amplitudes including all
overall coupling and averaging factors~\cite{Heinrich:2017kxx} -- and takes the general form,
\begin{equation}
{\cal V}_b = {\cal N} \, \frac{\alpha_s}{2\pi} \left[
\frac{1}{\epsilon^2} \, a \, {\cal B} + \frac{1}{\epsilon} \sum_{i \neq j} c_{ij} \, {\cal B}_{ij}
 + {\cal V}_{\rm fin} \right] \;, \qquad
{\cal N} = \frac{(4\pi)^\epsilon}{\Gamma(1-\epsilon)} \left( \frac{\mu^2}{Q^2} \right)^\epsilon \;,
\label{eq:Vfin}
\end{equation}
where ${\cal B}$ is the Born contribution.  From colour conservation, for this process we have
${\cal B}_{12} = {\cal B}_{21} = C_A {\cal B}$.  The coefficients of the pole terms are
determined by $a = -2 C_A$ and $c_{12} = c_{21} = -\beta_0/C_A - \log(\mu^2/\hat s)$, where
$\beta_0 = (11 C_A - 2n_f)/6$.\footnote{
This corrects the definition of $c_{12}$ and $c_{21}$ found in the published version of ref.~\cite{Heinrich:2017kxx};
the arXiv version has been updated.}
Finally, the value of ${\cal V}_{\rm fin}$ is provided for the scale $\mu_0=\sqrt{\hat s}/2$,
from which the result at an arbitrary scale $\mu$ can be found using,
\begin{equation}
{\cal V}_{\rm fin}(\mu) ={\cal V}_{\rm fin}(\mu_0) \cdot \frac{\alpha_s^2(\mu)}{\alpha_s^2(\mu_0)}
 + C_A {\cal B}(\mu) \left[ \log^2\left( \frac{\mu_0^2}{\hat s} \right)
  - \log^2\left( \frac{\mu^2}{\hat s} \right) \right]
\label{eq:Vfinrun}
\end{equation}
The virtual contribution is implemented in {\tt MCFM} by using eqs.~(\ref{eq:Vfin})
and~(\ref{eq:Vfinrun}), setting ${\cal N}=1$ since such an overall factor
is implicit in the rest of the code.

\subsection{Validation}

We first compare with the 14 and 100 TeV total cross section results presented in ref.~\cite{Borowka:2016ypz}.
These are obtained with $m=173$~GeV, $\mh=125$~GeV and the PDF set
{\tt PDF4LHC15\_nlo\_100\_pdfas} (for both LO and NLO calculations).  The top quark width is set to
zero.

In order to address issues of numerical stability,
we have implemented a rescue system in the calculation of the
real radiation corrections.  This compares the calculation of the matrix
elements at two phase-space points related by a rotation in order to
provide an estimate of the numerical accuracy.  If this suggests that
the result is not accurate to at least eight digits we switch on the
fly from double to quadruple precision and repeat the calculation of
the real emission matrix elements.  Although this rescue system
clearly requires two evaluations of the matrix elements, since they
are already very fast to compute this is not a great additional
burden.  With this in place we obtain the same integrated
cross-sections as when using {\tt OpenLoops}, but in a factor 50 less time.
To eliminate unnecessary further numerical instability,
we have also imposed a technical cut $p_T(HH)/\sqrt{\hat s} > 10^{-2}$ . We have checked that variation of this cut
in the range $10^{-6}$ to $ 10^{-2}$ makes a difference of less than one per mille in the total
NLO result.

\begin{table}
\begin{center}
\begin{tabular}{l|c|c|c}
$\sqrt{s} [TeV]$ & Calculation & LO [fb] & NLO [fb] \\ \hline
14   & {\tt MCFM}                  & $19.85^{+27.6\%}_{-20.5\%}$ & $32.91^{+13.6\%}_{-12.6\%}$ \\
        & Ref.~\cite{Borowka:2016ypz} & $19.85^{+27.6\%}_{-20.5\%}$ & $32.91^{+13.6\%}_{-12.6\%}$ \\ \hline
100  & {\tt MCFM}                  & $730.9^{+20.9\%}_{-15.9\%}$ & $1149^{+10.8\%}_{-10.0\%}$ \\
        & Ref.~\cite{Borowka:2016ypz} & $731.3^{+20.9\%}_{-15.9\%}$ & $1149^{+10.8\%}_{-10.0\%}$ \\
\end{tabular}
\caption{Validation of 14 and 100 TeV cross sections against the results of ref.~\cite{Borowka:2016ypz}.
The numerical uncertainty in the {\tt MCFM} results is beyond the last digit. The percentage deviations
correspond to estimates of uncertainty from 7-point variation of the scale according to the procedure
described in ref.~\cite{Borowka:2016ypz}.} 
\label{tab:xsecvalidation}
\end{center}
\end{table}

\begin{figure}[t]
\centering
\includegraphics[width=0.3\textwidth,angle=270]{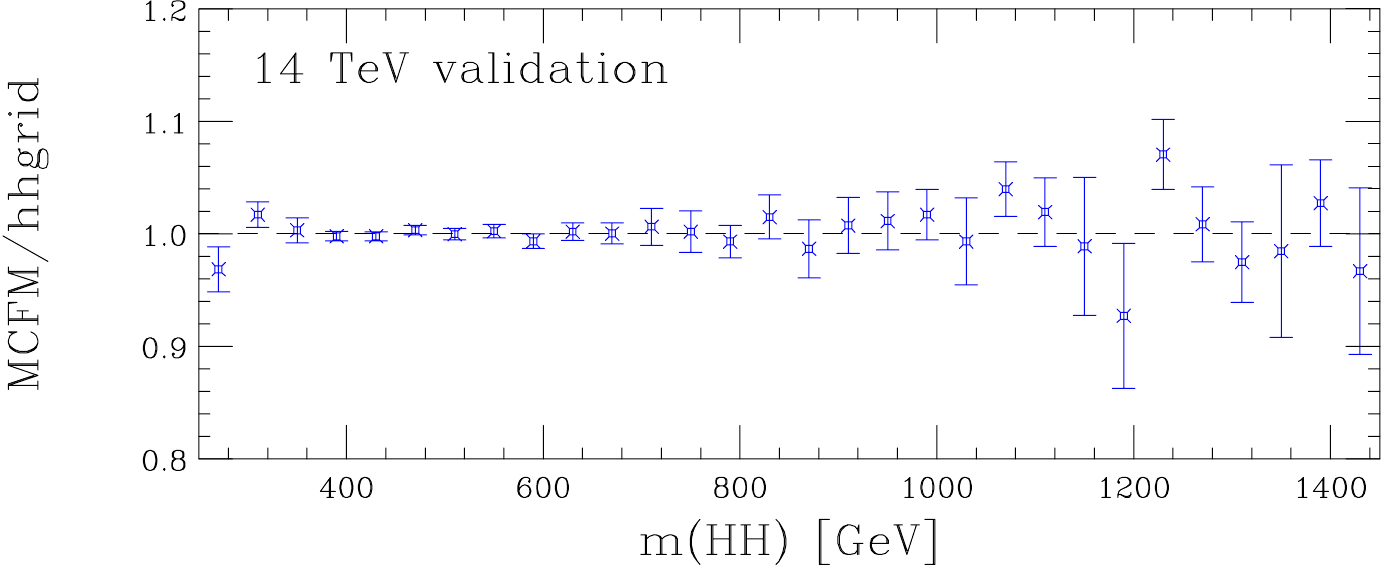}
\caption{Validation plot for the $m(HH)$ distribution, comparing against the {\tt hhgrid} result of ref.~\cite{Heinrich:2017kxx}.}
\label{fig:mHH}
\end{figure}

\begin{figure}[t]
\centering
\includegraphics[width=0.3\textwidth,angle=270]{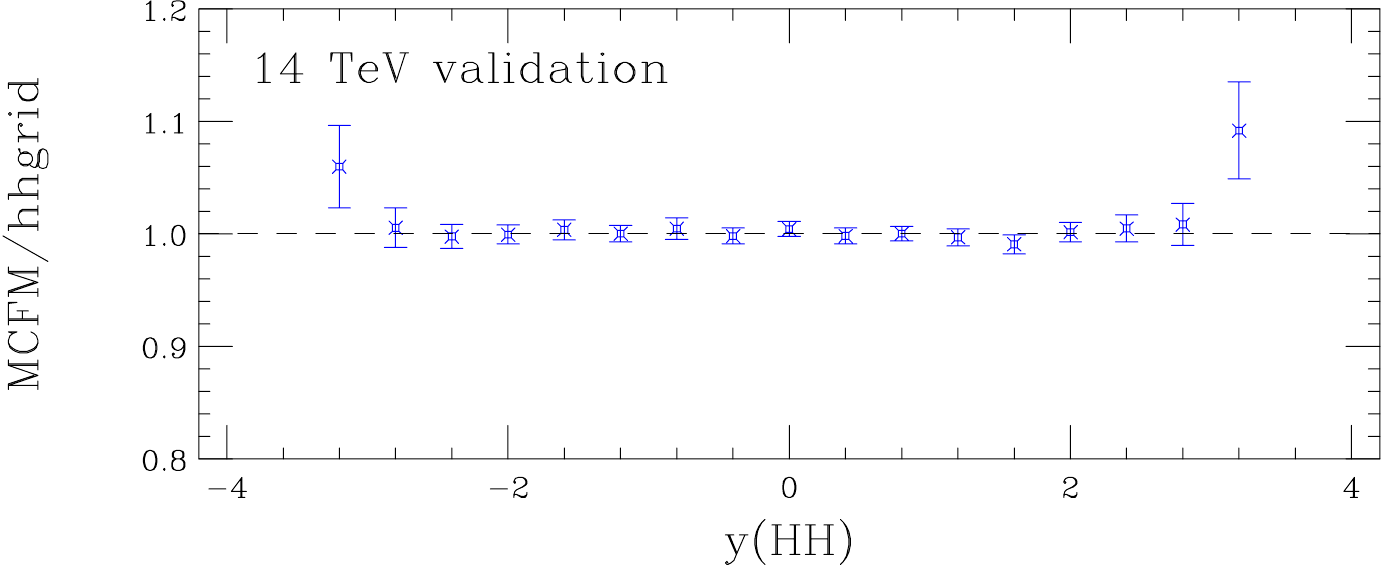}
\caption{Validation plot for the $y(HH)$ distribution, comparing against the {\tt hhgrid} result of ref.~\cite{Heinrich:2017kxx}.}
\label{fig:yHH}
\end{figure}

\begin{figure}[t]
\centering
\includegraphics[width=0.3\textwidth,angle=270]{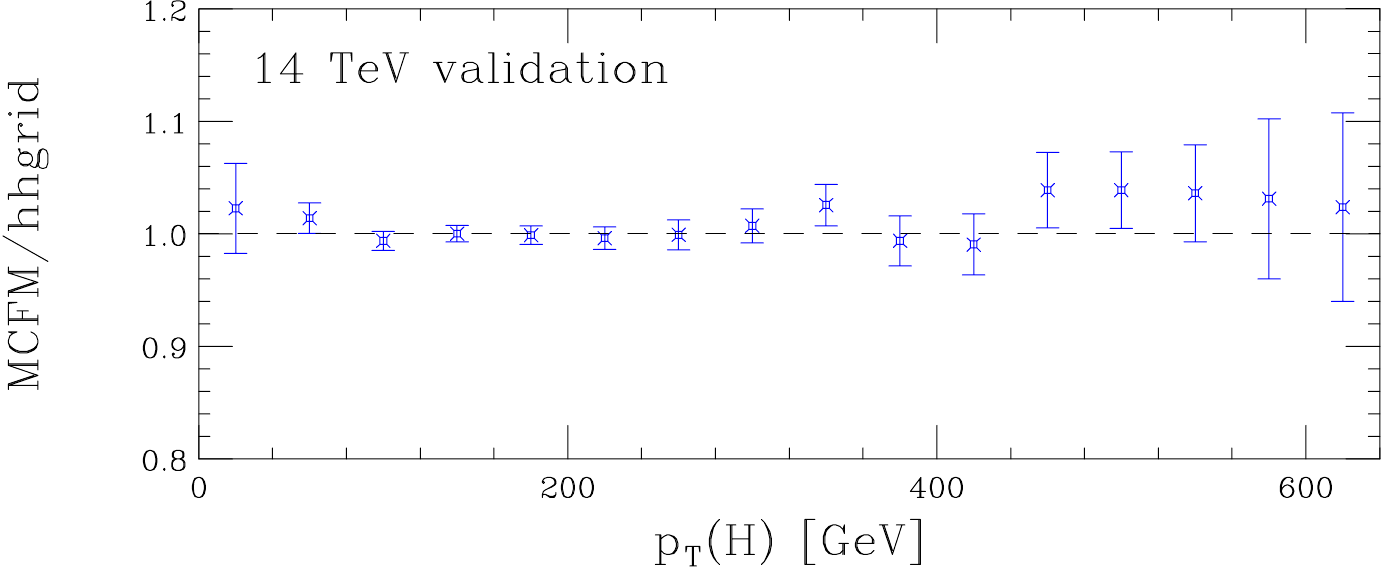}
\caption{Validation plot for the $p_T(H)$ (random Higgs boson) distribution, comparing against the {\tt hhgrid} result of ref.~\cite{Heinrich:2017kxx}.}
\label{fig:ptH}
\end{figure}

At the level of total cross sections, the perfect agreement between the NLO results from
the two codes is demonstrated in Table~\ref{tab:xsecvalidation}.
The {\tt hhgrid} package also provides 14 TeV validation data for the distributions of the Higgs boson pair
invariant mass and rapidity, as well as the transverse momentum of a random Higgs boson.
We compare these with the {\tt MCFM} results in figures~\ref{fig:mHH},~\ref{fig:yHH} and~\ref{fig:ptH},
which again demonstrate excellent agreement.

\subsection{Phenomenology}
With the fixed order Higgs pair production process implemented in {\tt MCFM} it is straightforward
to extend the existing framework to provide a resummed prediction for the
transverse momentum of the Higgs boson pair.  This is implemented using
the {\tt CuTe-MCFM} framework~\cite{Becher:2020ugp,Campbell:2022uzw};
we refer the reader to the original papers for more details.  For our
purposes it is important to note that for our matched resummed prediction,
which combines the NNLL result at small $q_T$ with the fixed order one
at high $q_T$, we use a transition function with parameter $x=q_T^2(HH)/m^2_{HH}$
and $x_{max}=0.1$.

\begin{figure}[t]
\centering
\includegraphics[width=0.45\textwidth,angle=270]{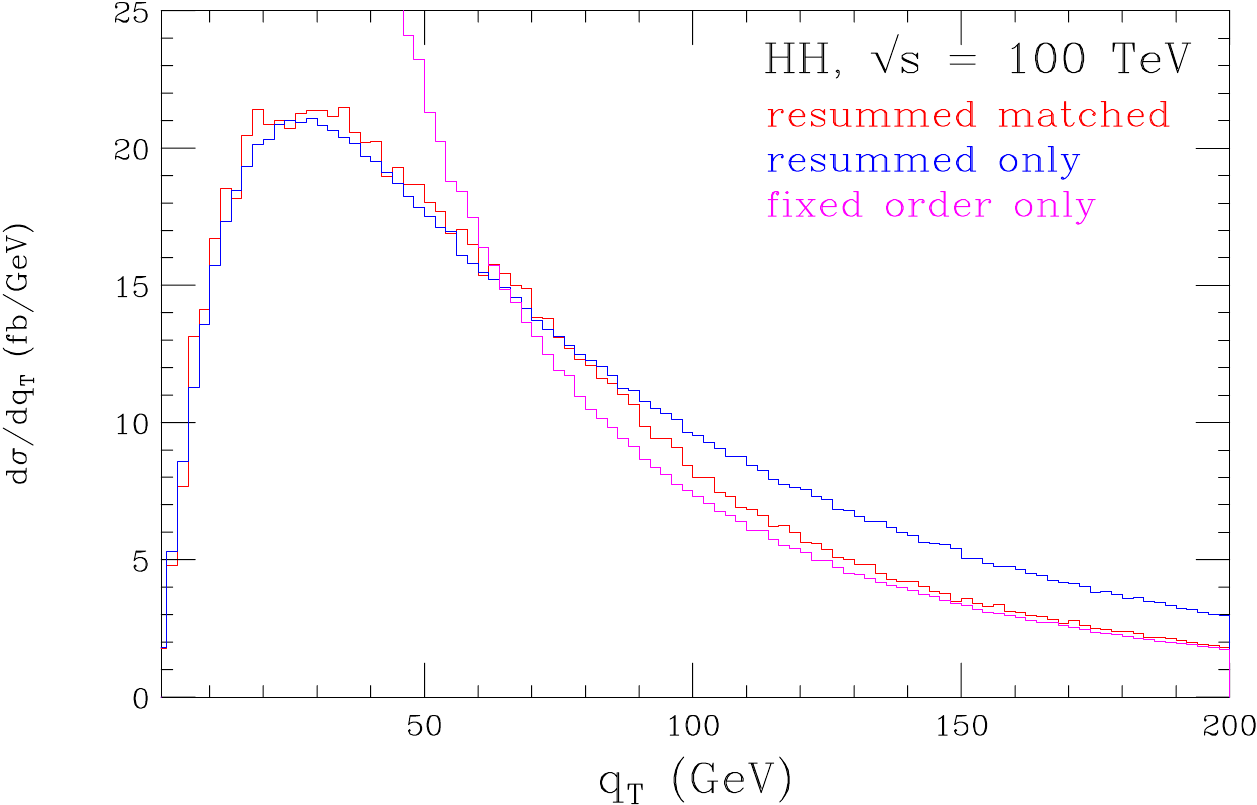}
\caption{The $q_T$ spectrum of the Higgs
boson pair at a $100$~TeV $pp$ collider.  The plot shows the
matched resummed prediction (red), compared with the pure resummed result (blue) and the
fixed order calculation (magenta).
}
\label{fig:ptHHresummed}
\end{figure}

As an example, in Fig.~\ref{fig:ptHHresummed} we show
the matched, resummed $q_T$ spectrum of the Higgs boson pair at
a $100$~TeV $pp$ collider, together
with the results obtained from pure fixed order and resummed calculations.
The fixed order result clearly diverges at small $q_T$ while the resummed
result ameliorates this behavior.  The matched resummed result smoothly
interpolates between the two calculations, transitioning to the fixed-order
result for $q_T$ around
$q_T^{max} = \sqrt{x_{max}} \, m_{HH} \sim 130$~GeV (using the fact that
the peak of the $m_{HH}$ distribution is around $400$~GeV). 
The matched result begins to differ substantially from the resummed result
a little before that, around $q_T \sim 90$~GeV.

\section{Conclusions}
This paper has addressed the calculation of the amplitude for a pair of Higgs bosons
in association with three partons at one-loop level. The calculation proceeded in two steps. First the
coefficients of the needed scalar integrals were calculated using both methods based on
the work of Passarino and Veltman, as well as more modern techniques based on generalized unitarity.
The initial results for the box coefficients obtained using unitarity were subsequently simplified
using the technique of analytic reconstruction. Compared to previous uses of this technique, new
strategies were introduced in order to handle particular features of this processes, in particular
the presence of two massive external particles and a massive particle circulating in the loop. 
This latter step, which yielded the simpler results for the box
coefficients and some of the triangle coefficients given in this paper,
also improved the speed of the numerical evaluation. The resulting code,
in combination with previous work~\cite{Heinrich:2017kxx} on the two loop corrections
to the Higgs boson pair + 2 parton
process, allows the fast evaluation of the next-to-leading order corrections to $pp \to HH$.
This calculation will be included in an upcoming release of the {\tt MCFM} code,
also providing machine-readable versions of the analytic amplitude results presented in this paper.

\section*{Acknowledgments}
We acknowledge useful discussions with Stephen Jones. RKE acknowledges
receipt of a Leverhulme Emeritus Fellowship from the Leverhulme Trust.
GDL’s work is supported in part by the U.K.~Royal Society through
Grant URF$\backslash$R1$\backslash$20109.
The work of J.M.C. is supported in part by the U.S. Department of Energy, Office of Science, Office of Advanced Scientific Computing Research, Scientific Discovery through Advanced Computing (SciDAC-5) program, grant “NeuCol”.
This manuscript has been
authored by Fermi Research Alliance, LLC under Contract
No. DE-AC02-07CH11359 with the U.S. Department of Energy, Office of
Science, Office of High Energy Physics.  This research used resources
of the National Energy Research Scientific Computing Center (NERSC), a
U.S.\ Department of Energy Office of Science User Facility, under
NERSC award HEP-ERCAP0023824.

\appendix
\section{Spinor algebra}
\label{spinorsection}
All results are presented using the
standard notation for the kinematic invariants of the process,
\begin{equation}
s_{ij} = (p_i+p_j)^2 \, ,
s_{ijk} = (p_i+p_j+p_k)^2 \, ,
s_{ijkl} = (p_i+p_j+p_k+p_l)^2 \,.
\end{equation}
and the Gram determinant,
\begin{equation} \label{Delta3eqn}
\DeltaThree(i,j,k,l) =(s_{ijkl}-s_{ij}-s_{kl})^2-4 s_{ij} s_{kl}  \, .
\end{equation}
In the case where momentum $j$ is not lightlike, we put the corresponding subscript in boldface, e.g.~$s_{i {\bm j}}$.
We express the amplitudes in terms of spinor products defined as,
\begin{equation}
\label{Spinor_products1}
\spa i.j=\bar{u}_-(p_i) u_+(p_j), \;\;\;
\spb i.j=\bar{u}_+(p_i) u_-(p_j), \;\;\;
\spa i.j \spb j.i = 2 p_i \cdot p_j,\;\;\;
\end{equation}
and we further define the spinor sandwiches for massless momenta $j$ and $k$,
\begin{eqnarray}
\label{Spinor_products2}
\spab{i}.{(j+k)}.{l} &=& \spa{i}.{j} \spb{j}.{l} +\spa{i}.{k} \spb{k}.{l} \nn \\
\spba{i}.{(j+k)}.{l} &=& \spb{i}.{j} \spa{j}.{l} +\spb{i}.{k} \spa{k}.{l}
\end{eqnarray}
The spinor sandwich with momentum $k$ not lightlike is distinguished by putting the momentum
$k$ in boldface, e.g.~$\spab{i}.{\bm k}.{l}$. 

In the Weyl representation for the Dirac gamma matrices, following the conventions of
ref.~\cite{Peskin:1995ev}, we have
\begin{equation} 
\slsh{p}=\gamma^0 p^0 -\gamma^1 p^1 -\gamma^2 p^2 -\gamma^3 p^3
= \left(\begin{matrix}
0& 0 & p^- & -p^1+ip^2\cr
0& 0 & -p^1-i p^2 & p^+\cr
p^+ & p^1-ip^2& 0 & 0 \cr
p^1+i p^2 & p^- & 0 & 0 
\end{matrix}
\right)\,.
\end{equation}
The massless spinors solutions of Dirac equation are
\beq \label{eq:explicitspinorL}
u_-(p) =
\left[ \begin{matrix} (-p^1+ip^2)/\sqrt{p^+} \cr
    \sqrt{p^+}\cr
     0\cr
     0 \cr  \end{matrix}\right]
=
\left[ \begin{matrix} -\sqrt{p^-} e^{- i\varphi_p}\cr
    \sqrt{p^+}\cr
     0\cr
     0 \cr  \end{matrix}\right] , 
\eeq
and
\beq \label{eq:explicitspinorR}
u_+(p) =
  \left[ \begin{matrix} 0 \cr
                  0 \cr
                  \sqrt{p^+} \cr
                  (p^1+ip^2)/\sqrt{p^+} \cr
                 \end{matrix}\right] 
=
  \left[ \begin{matrix} 0 \cr
                  0 \cr
                  \sqrt{p^+} \cr
                  \sqrt{p^-}  e^{i\varphi_p}
                 \end{matrix}\right] \,,
\eeq
where 
\beq \label{eq:phasekdef}
e^{\pm i\varphi_p}\ \equiv\ 
  \frac{ p^1 \pm ip^2 }{ \sqrt{(p^1)^2+(p^2)^2} }
\ =\  \frac{ p^1 \pm ip^2 }{ \sqrt{p^+p^-} }\ ,
\qquad p^\pm\ =\ p^0 \pm p^3.  
\eeq
In this representation the Dirac conjugate spinors are
\beq
\label{eq:explicitspinorconjg}
\overline{u}_-(p) \equiv  u_-^\dagger(p) \gamma^0 =
  \left[ 0, 0, -\sqrt{p^-} e^{i\varphi_p} ,\sqrt{p^+} \right] 
\eeq
\beq
\overline{u}_+(p) \equiv  u_+^\dagger(p) \gamma^0 =
\left[\sqrt{p^+}, \sqrt{p^-} e^{-i\varphi_p},0,0 \right] 
\eeq
More complicated spinor products, follow in an obvious way,
\beq
\spab{i}.{{\bf k}}.j=\bar{u}_{-}(p_i) \, \slsh{{\bf k}}\, u_{-}(p_j)\,, \quad\quad
\spaba{i}.{{\bf k}}.{{\bf l}}.j=\bar{u}_{-}(p_i) \, \slsh{{\bf k}}\,\slsh{{\bf l}}\; u_{+}(p_j) \,,
\eeq
where ${\bf k}$ and ${\bf l}$ are the momenta of non lightlike particles.
 \section{Loop integral definitions}
\label{Integrals}
We work in the Bjorken-Drell metric so that
$l^2=l_0^2-l_1^2-l_2^2-l_3^2$. The affine momenta $q_i$
are given by sums of the external momenta, $p_i$, 
where $q_n\equiv \sum_{i=1}^n p_i$ and $q_0 = 0$.
The propagator denominators are defined as $ d_i=(l+q_i)^2-m^2+i\varepsilon$.
The definition of the relevant scalar integrals is as follows,
\begin{eqnarray} \label{eq:scalarintegrals}
&& B_0(p_1;m)  =
 \frac{\mu^{4-D}}{i \pi^{\frac{D}{2}}\cG}\int \,
 \frac{d^D l} {d_0 \; d_1}\, , \nn \\
&& C_0(p_1,p_2;m)  =
\frac{1}{i \pi^{2}}
\int \, 
 \frac{d^4 l}{d_0 \; d_1 \; d_2}\, ,\nn \\
&&D_0(p_1,p_2,p_3;m)
= \frac{1}{i \pi^{2}} \int \, \frac{d^4 l} {d_0 \; d_1 \; d_2\; d_3}\, ,\nn \\
&&E_0(p_1,p_2,p_3,p_4;m)
= \frac{1}{i \pi^{2}} \int \,\frac{d^4 l}{d_0 \; d_1 \; d_2\; d_3\; d_4}\, .
\end{eqnarray}
For the purposes of this paper we take the masses in the
propagators to be real.  Near four dimensions we use $D=4-2 \e$.  (For
clarity the small imaginary part which fixes the analytic
continuations is specified by $+i\,\varepsilon$).  $\mu$ is a scale introduced so that the integrals
preserve their natural dimensions, despite excursions away from $D=4$.
We have removed the overall constant which occurs in $D$-dimensional integrals,
\beq
\cG\equiv\frac{\Gamma^2(1-\e)\Gamma(1+\e)}{\Gamma(1-2\e)} = 
\frac{1}{\Gamma(1-\e)} +{\cal O}(\e^3) =
1-\e \gamma+\e^2\Big[\frac{\gamma^2}{2}-\frac{\pi^2}{12}\Big]
+{\cal O}(\e^3)\,.
\eeq
\section{Spinor decompositions}\label{sec:spinor_decompositions}

We supplement the primary decompositions presented in
ref.~\cite[Section 3.3.1]{Campbell:2022qpq} with the following new
decomposition
\begin{align} \label{eq:primary_decomposition_spbab1.4.5_spbab2.4.5.2}
  \Big\langle \spbab1.(4+5).(6+7).1, \spbab2.(4+5).(6+7).2 \Big\rangle = & \nn \\
  & \kern-50mm \Big\langle \spb1.2, \spb1.3, \spb2.3 \Big\rangle \; \cap \; \Big\langle \spb1.2, \spab3.(4+5).2, \spab3.(4+5).1 \Big\rangle \; \cap \\
  & \kern-50mm \Big\langle  \spbab1.(4+5).(6+7).1, \spbab2.(4+5).(6+7).2, \nn \\
  & \kern-30mm |(1+3)|(4+5)|2][1|+|2][1|(4+5)|(2+3)|-|2]\spab3.(4+5).3[1| \Big\rangle \, , \nn 
\end{align}
where it appears that the last ideal on the right-hand side may
require further decompositions, even if this form suffices for the
current discussion. The generator with two open indices was obtained
by fitting a covariant ansatz to numerical evaluations obtained from a
component expression obtained from \texttt{Singular}, after
quotienting the left-hand side ideal by the first two prime ideals on
the right-hand side. The equality holds in $R_7$ without imposing the
additional constraint $s_{45}=s_{67}$. Changing the ring, e.g.~to
$\mathcal{R}\kern-2.90mm\mathcal{R}_5$, causes non-trivial
modification to the decomposition.

The decomposition of
eq.~\eqref{eq:primary_decomposition_spbab1.4.5_spbab2.4.5.2} can be
understood in light of the following identity
\begin{equation}\label{eq:explicitfacotrizationnewideal}
  \begin{gathered}
    \kern-25mm |1]\spbab2.(4+5).(6+7).2[1|-|2]\spbab1.(4+5).(6+7).1[2| =  \\
    \kern+25mm  - \spb1.2 \Big( |(1+3)|(4+5)|2][1|+|2][1|(4+5)|(2+3)|-|2][3|(4+5)|3][1| \Big)
  \end{gathered}
\end{equation}
where on the left-hand side we clearly have a member of the maximal
codimension ideal being decomposed in
eq.~\eqref{eq:primary_decomposition_spbab1.4.5_spbab2.4.5.2}, while on
the right-hand side we have a polynomial with two factors. This
manifestly shows that a non-trivial primary decomposition is needed,
and identifies the two factors as generators of ideals in the
decomposition.

We can see this decomposition reflected in the structure of the
coefficient $c_{4\x123}^{(2)}(1^-,2^-,3^+)$ from
eq.~\eqref{eq:c4x123mmpxm2}. In order to fully separate the poles
$\spbab1.\four.\five.1$ and $\spbab2.\four.\five.2$ into separate
fractions, it is necessary to introduce a spurious second power of the
pole $\spb1.2$. Without it, the common numerator does not vanish on
all branches of
\eqref{eq:primary_decomposition_spbab1.4.5_spbab2.4.5.2}, thus
preventing the partial fraction decomposition by Hilbert's
Nullstellensatz.

Alternatively, it is possible to avoid introducing the spurious double pole, if $\spbab1.\four.\five.1$ and $\spbab2.\four.\five.2$ are kept in the same denominator
\begin{gather}
c^{(2)}_{4\x123}(1^-,2^-,3^+) =
\Big\{2\*\frac{\spb1.3^3\*s_{\four\five}(s_{\four\five}-4\*\mh^2)}{\spb1.2\*\spb2.3\*\spbab1.\four.\five.1^2}
+4\frac{\spb1.3 \spbab3.\four.\five.3}{\spb1.2\spb2.3\spbab1.\four.\five.1}-2\frac{\spb1.3\spab2.\four-\five.3}{\spb1.2\spbab1.\four.\five.1} \Big\} -\Big\{ 1\leftrightarrow 2 \Big\} \nn \\
- \frac{1}{2} \frac{\tr(3-1-2|\four-\five) (\spb1.3\spb2.3\tr(3-1-2|\four-\five) - \spb1.2\spbab3.(\four-\five).(1-2).3)}{\spb1.2\spbab1.\four.\five.1\spbab2.\four.\five.2} \label{eq:c4x123mmpxm2-alternative} \, .
\end{gather}
This form also manifests the symmetry under $\bold 4 \leftrightarrow \bold 5$ term by term.
The numerator of the last fraction is now a contraction of the covariant generator in \eqref{eq:primary_decomposition_spbab1.4.5_spbab2.4.5.2}, 
\begin{equation}
  \frac{1}{2} \spb1.2 (\spb1.3\spb2.3\tr(3-1-2|\bold 4-\bold 5) - \spb1.2\spbab3.(\bold 4- \bold 5).(1-2).3) = \spb1.3^2\spbab2.\bold 4.\bold 5.2-\spb2.3^2\spbab1.\bold 4.\bold 5.1 \, ,
\end{equation}
where for convenience we have written this as $[3| \times $
  \eqref{eq:explicitfacotrizationnewideal} $\times |3]$.  As before the trace is
understood as being of rank-two spinors:
$(3-1-2)_{\alpha\dot\alpha}(\four-\five)^{\dot\alpha\alpha}$.
 \section{Box manipulations and spurious singularities at infinity}\label{sec:boxmanipulation}

In this appendix we analyse the singularity structure of one of the
box coefficients, $d_{1\x2\x3}(1^-,2^-,3^+)$, which was presented
eq.~\eqref{eq:d123mmpgggHH}. That form with the effective pentagon
involves a spurious single pole in $\trfive$, as well a degree two
(instead of one) pole in $[12]$, a degree one (instead of two) zero in
$\spa1.2$, etc. A form with more manifest analytic properties is,
\begin{eqnarray}\label{eq:d1x2x3mmpfull}
  d_{1\x2\x3}(1^-,2^-,3^+) &=& \frac{
    \begin{aligned}
        \spa1.2^2\spa2.3[23]m^2&\Big([3|\four|\five|3](\spa1.3[3|\four|\five|1]-[13]\spaba1.\five.\four.3) \\
        &+\spa1.2[23](s_{123}-2s_{12}-2\mh^2+8m^2))[3|\four|\five|1]\Big)
    \end{aligned}
  }{8(-\spa1.2[12]\spa2.3[23]\spaba1.\five.\four.3[3|\four|\five|1]+m^2\trfive^2)} + (\four \leftrightarrow \five) \nn \\
  && +\frac{\begin{aligned}
      &\spa1.2^3[13]\spa2.3[23]^2m^4\trfive(\spaba1.\four.\five.3[3|\five|\four|1]-(\four \leftrightarrow \five))(s_{123}-2s_{12}-2\mh^2+8m^2)
    \end{aligned}
  }{4(-\spa1.2[12]\spa2.3[23]\spaba1.\five.\four.3[3|\four|\five|1]+m^2\trfive^2)\times(\four \leftrightarrow \five)} \nn \\
  && + m^2\frac{\spa1.2^2[23]}{2[12]\spa1.3} \, ,
\end{eqnarray}
where the first line is a doublet and the latter two lines are
singlets under the exchange of the Higgs bosons,
$\four\leftrightarrow\five$. In eq.~\ref{eq:d1x2x3mmpfull}
this is shown explicitly. Except for the
contribution in the last line of eq.~\eqref{eq:d1x2x3mmpfull}, the
only singularities are $|S^{1\x2\x3\x4}|$ and
$|S^{1\x2\x3\x4}|(\four\leftrightarrow\five)$, as introduced in
eq.~(\ref{Sdef2}).

Let us now analyse the large $m$ limit. As discussed in section
\ref{sec:prjectivem}, we consider a projective space in $m$, rather
than an affine space. This box coefficient diverges linearly in
$m^2$. In the limit, it reads,
\begin{eqnarray}\label{eq:d1x2x3mmplargem}
  \lim_{m\rightarrow \infty}d_{1\x2\x3}(1^-,2^-,3^+) &=&
  + m^2\frac{\spa1.2^2[23]}{2[12]\spa1.3}
  -m^2\frac{\spa1.2^3[13]\spa2.3[23]^2(s_{123}-2\mh^2)}{\trfive^2} \nn
  \\ &&
  +2m^2\frac{\spa1.2^3[13]\spa2.3[23]^2(\spaba1.\four.\five.3 [3|\five|\four|1]-(\four
    \leftrightarrow \five))}{\trfive^3} \, .
\end{eqnarray}
This form can be read out from eq.~\eqref{eq:d1x2x3mmpfull}, using the
following identity,
\begin{equation}
  [1|\four|\five|3]+[1|\five|\four|3] = [13](s_{123}-2\mh^2) \, .
\end{equation}

We can now understand why the two $|S|$ denominators in the general
$m$ expression of eq.~(\ref{eq:d1x2x3mmpfull}) cannot be separated
without introducing a spurious pole in $\trfive\rightarrow 0$ or
$m\rightarrow \infty$. An ansatz where the $|S|$ denominators are
separated would read, schematically,
\begin{equation}
  d_{1\x2\x3} \sim \sum_{\text{terms}} \frac{m^\alpha\trfive^\beta}{|S|}
\end{equation} 
On the irreducible codimension-two projective variety
$V\big(\big\langle \trfive, m^{-1} \big\rangle\big)$ we must have a
pole of degree five, as manifest in the second line in
eq.~(\ref{eq:d1x2x3mmplargem}), thus we have the constraint,
\begin{equation}
  \alpha - \beta = 5 \, .
\end{equation}
We impose the equality because at least one of the terms must saturate
the limit.  On the irreducible codimension-one projective variety
$V\big(\big\langle m^{-1} \big\rangle\big)$ we have a double pole,
i.e.
\begin{equation}
  \alpha \leq 4 \, ,
\end{equation}
since $|S|$ goes quadratically. We allow the inequality since not all
terms need to saturate this limit. On the irreducible codimension-one
variety $V\big(\big\langle \trfive \big\rangle\big)$ the coefficient
is regular,
\begin{equation}
  \beta \geq 0 \, .
\end{equation}

In the $\alpha-\beta$ plane the line $\alpha - \beta = 5$ does not
intersect the semi-infinite region defined by $\alpha \leq 4 \, ,
\beta \geq 0$. Therefore, no simultaneous solution exists to these
three constraints.

Consider one last form for this coefficient,
\begin{eqnarray}
  d_{1\x2\x3}(1^-,2^-,3^+) &=&
  \bigg\{ \frac{\spa1.2^2[23]m^2}{8(-\spa1.2[12]\spa2.3[23]\spaba1.\five.\four.3[3|\four|\five|1]+m^2\trfive^2)} \times \nn \\
    && \; \Big[ (s_{123}-2s_{12}-2\mh^2+8m^2)(-2m^2\trfive\frac{[13]}{[12]}+\spa1.2\spa2.3[23][3|\four|\five|1]) \nn \\
    && \; +\spa2.3[3|\four|\five|3](\spa1.3[3|\four|\five|1]-[13]\spaba1.\five.\four.3) \Big] \bigg\} + \bigg\{\four \leftrightarrow \five \bigg\} \nn \\
  && + m^2\frac{\spa1.2^2[23]}{2[12]\spa1.3}
\end{eqnarray}
The above curly bracket goes like $m^4$ in the large $m$ limit (the
numerator goes like $m^6$, the denominator as $m^2$), while we have
already shown that this box coefficient only scales as $m^2$. In terms
of $\alpha$ and $\beta$, we have $\alpha = 6$ and $\beta = 1$. This
form has a spurious pole at infinity, meaning it appears in the
numerator rather than the denominator.
 \bibliography{HiggsPair}       
\bibliographystyle{JHEP}       
\end{document}